\documentclass[11pt,a4paper]{article}
\usepackage{jheppubmod}
\usepackage{tikz-feynman}
\tikzfeynmanset{compat=1.0.0}
\usepackage{graphicx}
\usepackage{amssymb,mathtools}
\usepackage{amsmath}
\usepackage{epstopdf}
\usepackage{hyperref}
\usepackage{color}
\usepackage{mathrsfs}
\usepackage{cite}
 
\newcommand{\half}{ \frac{1}{2}}

\def\dm2{{d-2 \over 2}}
\def \half{\frac{1}{2}}
\def\dflat{\mathcal{D}^{\flat}}
\def\A[#1]{A^{(#1)}}
\def\d[#1]{\Delta_{#1}}
\def\D[#1]{\Delta^{(#1)}}
\def\Dv[#1]{\Delta^{[#1]}}
\def\gflata[#1,#2][#3,#4]{\mathcal{G}^{\flat}_{\d[#1],\ldots,\d[#2]}(x_{#3},\ldots,x_{#4})}
\def\dflata[#1,#2]{\mathcal{D}^{\flat}_{\d[#1],\ldots,\d[#2]}(\{r_i\})}
\def\dcfta[#1,#2]{\mathcal{D}^{cft}_{\d[#1],\ldots,\d[#2]}(\{s_i\})}
\def\gcfta[#1,#2][#3,#4]{\mathcal{G}^{cft}_{\d[#1],\ldots,\d[#2]}(x_{#3},\ldots,x_{#4})}
\def\gflat{\mathcal{G}^{\flat}}
\def\dcft{\mathcal{D}^{cft}}
\def\gcft{\mathcal{G}^{cft}}

\def\sumi[#1]{\sum_{i=1}^{[#1]}}

\def\iflat{\mathcal{I}^{\flat}}
\def\iflata[#1,#2][#3,#4]{\mathcal{I}^{\flat}_{\d[#1],\ldots,\d[#2]}(x_{#3},\ldots,x_{#4})}

\def\gflate[#1,#2][#3,#4]{\mathcal{G}^{\flat}_{\d[1],\ldots,\d[#1];\d[#2]}(x_{#3},\ldots,x_{#4})}
\def\gflate4[#1]{\mathcal{G}^{\flat}_{\d[1],\ldots,\d[4];#1}(x_{1},\ldots,x_{4})}
\def\dflate4{\mathcal{D}^{\flat}_{\d[1],\ldots,\d[4];\d[0]}(r_{1},\ldots,r_{5})}

\title{Conformality of non-conformal correlators}
\author{Siddharth G. Prabhu}
\affiliation{Department of Theoretical Physics, Tata Institute
	of Fundamental Research, Homi Bhabha Rd, \\Mumbai 400005, India}
\emailAdd{siddharth.g.prabhu@gmail.com}

\date{}                             
\begin{document}
	
	\abstract{We show that position space correlators of a Poincare invariant quantum field theory can be recast in terms of conformally invariant correlators, in other words, as functions of conformal cross ratios. In particular, we show that correlators of massless fields in flat spacetimes with $n-$point contact interactions can be expressed as position space soft limits of conformally invariant correlators with $(n+1)-$point interactions. We show that this correspondence applies at the level of every Feynman diagram that appears in the perturbative expansion of the correlators in the respective coupling constants. We apply this method to find exact answers for some Feynman diagrams including several loop examples. We also show that the analogous correlators for massive fields can be expressed as infinite sums of conformal correlators.}
	
	\maketitle
	
\section{Introduction}

The degree of complexity involved in the analysis of physical systems decreases considerably with increasing symmetry of the system. While spacetime processes are naturally Poincare invariant, for some applications it turns out to be useful to think about the more symmetric conformally invariant systems. The presence of the additional conformal symmetries lends several advantages including the practical one of making computations simpler. In this article, we show that correlation functions with massless scalar fields in flat spacetimes can be cast as conformally invariant correlators, by which we mean correlators that only depend on conformal cross ratios. We also show that correlators of massive scalar fields can be recast as infinite sums of such conformally invariant correlators. 

Consider a quantum field theory of massless fields in flat spacetime with an $n-$point non-derivative contact interaction. We show that all the correlators of this theory can be reproduced by conformal correlators corresponding to an $(n+1)$-point interaction. This interaction is between the original $n$ fields of the $n-$point contact interaction together with an extra field introduced so that the scaling dimensions associated with the $(n+1)-$ point interaction add up to the spacetime dimension. This latter \emph{conformal constraint} ensures that the interaction term is marginal. The correspondence between correlators works at the level of every diagram in their perturbative expansions in the respective coupling constants. 

Suppose one is interested in the contribution to a $m-$point correlator in the massless theory given by a Feynman diagram which has $V$ number of vertices. We derive the following procedure to obtain this contribution. At every vertex in the diagram we are interested in, we add an extra external field such the conformal constraint is satisfied at that vertex, thus obtaining a diagram that contributes to a conformal correlator. We take every spacetime point corresponding to each extra field added to infinity. In this \emph{position space soft limit}, the conformal answer factors into a product of soft factors (one for each extra field added) and the contribution to the $m-$point correlator one is interested in. 

We now describe this in some more detail. We work in $d$-dimensional flat spacetime. Let us start by describing the structure of contact contributions to the respective correlators. The contact contribution to the conformal $n$-point  function for fields $\phi_1(x_1),\ldots,\phi_n(x_n)$ with corresponding scaling dimensions $\d[1],\ldots,\d[n]$, satisfying the conformal constraint $\sum_{i=1}^{n} \d[i]=d$, has the form
\begin{equation}\label{gcft}
	\gcfta[1,n][1,n]= f(\{s_{ij}\}) \dcfta[1,n],
\end{equation}
where $f$ is a simple polynomial function of the distances between insertions $s_{ij}$, and $\dcft$ is a non-trivial function of the conformally invariant cross ratios $s_i$. 
Such conformally invariant correlators were first considered in \cite{Symanzik:1972wj} who provided a general procedure to compute them. They also arise when considering AdS Witten diagrams, and our $\dcft$ functions are denoted as $\mathcal{D}$ functions in that context \cite{DHoker:1999kzh}\footnote{We will denote them as $\dcft$ functions to distinguish them from their counterparts in the massless field theory we are interested in, and which we will introduce shortly.}.

Now, let us look at the structure of the $n-$point correlator in a massless field theory. Having no intrinsic scale, such a theory also enjoys scaling invariance. This means that in place of the conformal cross ratios, the spacetime invariants are instead simple ratios of spacetime distances of the form $r_i=\frac{s_{kl}}{s_{mn}}$. We can express the $n-$point contact contribution as

\begin{equation} \label{gflat}
	\gflata[1,n][1,n]= h(\{s_{ij}\}) \dflata[1,n],
\end{equation}
where $h$ is another simple function of the distances $s_{ij}$, and the non-trivial part of the correlator is contained in the $\dflat$ functions, introduced in \cite{Pibv}. Our central observation is the following relation between the $\dflat$ functions and their conformal counterparts:

\begin{equation}\label{dequality}
	{\mathcal {D}}^{cft}_{\d[1],\ldots,\d[n+1]}(\{s_i\})=\dflat_{\d[1],\ldots,\d[n]}(\{r_i\})
\end{equation}
under the identification $s_i=r_i$.

First, let us check that the counting of parameters on each side of \eqref{dequality} matches. The functions depend on scaling dimensions and spacetime invariants. As we have already noted, the conformally invariant correlation functions obey the constraint
\begin{equation}\label{confconstraint}
	\sum_i \d[i] =d
\end{equation}
which means that the number of scaling dimensions on each side matches.

Next, we count the number of independent pieces of spacetime data i.e. invariants of the spacetime coordinates on each side. For the massless field theory, we can fix two points completely. This is because we can use translations to fix the first point (say at the origin), rotations around the first point to bring a second point to a specified line, and scaling to fix its distance from the origin. Specifying a third point needs two numbers. Continuing in this manner, we need $(n-1)$ numbers to specify the $n^{th}$ point as long as $n-1\leq d$. Totally then, we need to specify $2+3+\ldots+n-1=n(n-1)/2-1$ numbers to fix all $n$ points. This number is the same as the number of independent simple ratios that can be formed from $n$ spacetime points, the latter being ${n \choose 2} -1=n(n-1)/2-1=(n+1)(n+1-3)/2$. On the other hand, if $n-1>d$, then to specify the first $(d+1)$ points, we need $1+2+\ldots+d=d(d+1)/2$ numbers, and to specify the last $(n-d-1)$ points, we need $d$ numbers each, for a total of $d(d+1)/2+(n-d-1)d=(n+1)d-1/2(d+1)(d+2)$ numbers to specify all $n$ points. 

To summarize, 
\begin{equation}
	\text{No. of simple ratios of n points}=\begin{cases}
		n(n-1)/2-1 & \ n-1 \leq d  \\
		 (n+1)d-1/2(d+1)(d+2) & \ n-1 > d 
	\end{cases}
\end{equation}

We can perform a similar counting as above in a conformal theory to find the well known result:  the number of independent cross ratios that can be formed from $n$ spacetime points is $\frac{n(n-3)}{2}$ when $n\leq (d+2)$, whereas for $n>d+2$, the number is $nd-\half(d+1)(d+2)$.

Hence, in each case, we see that the amount of independent spacetime data contained in  $n$ points in a Poincare invariant theory is the same as the amount of independent spacetime data contained in $(n+1)$ points in a conformally invariant theory. An easy way to see this is that using conformal invariance, an additional point can be fixed when compared to the Poincare invariant massless theory. In this way, we see that the number of independent parameters (scaling dimensions and spacetime coordinates) that the two sides of \eqref{dequality} and the corresponding correlation functions depend on, are equal to each other. 

The claim, as posed in \eqref{dequality} requires a formal identification of the arguments of the $\dcft$ function, the cross ratios of $(n+1)$ points  with the simple ratios of $n$ points that are the arguments of the $\dflat$ function. We provide a geometric realization of this identification. The geometric realization, which we refer to as the circumcentric configuration, is the following: the $(n+1)^{th}$ insertion is placed at the centre of the sphere that circumscribes the other $n$ insertions. 

In this circumcentric configuration, the conformal cross ratios involving the $(n+1)$ points become identical to the simple ratios formed by the $n$ points lying on the sphere (say of radius $r$). If we now perform a scaling by $1/r$, we reach the configuration with the $(n+1)^{th}$ point at the centre of a unit radius sphere on which the other $n$ points lie. In this unit circumcentric configuration, we will find that the dependence on distances of the two prefactors in Eqns.\eqref{gcft}  and \eqref{gflat} match, so that we have the following relation\footnote{We use the notation $\D[n]=\sum_{i=1}^{n} \d[i]$. The overall factor $\frac{1}{\Gamma(d-\D[n])}=\frac{1}{\Gamma(\d[n+1])}$ in \eqref{circconfig} comes about because of the one extra propagator in the $(n+1)$-point conformal correlator as compared to the $n$-point correlator, as is evident from the normalization of the propagator given in \eqref{prop}.}.
\begin{equation}\label{circconfig}
	\gflata[1,n][1,n]= \frac{1}{\Gamma(d-\D[n])} \gcft_{\d[1],\ldots,\d[n],d-\Delta^{(n)}}(x_1,\ldots,x_{n+1})|_{s_{1(n+1)}=\ldots=s_{n(n+1)}=1}
\end{equation}

The above result lends itself naturally to the interpretation of a \emph{soft limit} in position space. We define the soft limit of a position space conformal correlator as the limit in which an insertion is taken to infinity. This is related by an inversion to the unit circumcentric configuration. In momentum space, in the soft limit, scattering amplitudes factorize into the product of a soft factor and a scattering amplitude with one lesser insertion. In analogy with this, we shall see that in the position space soft limit, the conformal $(n+1)$-point correlation function factorizes into the product of a soft factor ${\cal{S}}(x_1,\ldots,x_{n+1})$ and a $n$-point massless correlation function,
\begin{equation}\label{contactfact}
	\lim_{x_{n+1}^2\to\infty} \gcft_{\d[1],\ldots,\d[n],d-\Delta^{(n)}}(x_1,\ldots,x_{n+1})={\cal{S}}(x_1,\ldots,x_{n+1}) \gflata[1,n][1,n]
\end{equation}
One way of making this result seem plausible is the following observation. Consider any configuration of points in a conformal theory with one point at infinity, and look for symmetries whose actions continue to keep the point at infinity. A special conformal transformation brings the point at infinity to a finite point. Hence, the class of conformal transformations that leaves the soft limit invariant does not include the special conformal transformation. In fact, in the soft limit, the remaining symmetries of the correlation function are rotations, translations and scalings. It is plausible, then, that the resulting object is a correlation function in a Poincare invariant theory of massless fields.

In momentum space, the soft factor blows up in the soft limit. It is natural, then, that the position space soft factor ${\cal S}(x_1,\ldots,x_n)$ vanishes in the soft limit.
We can write the inverse of Eq.\eqref{contactfact} to find the flat space correlator as a soft limit of the conformal correlator via
\begin{equation}\label{flatfromsoft}
	\gflata[1,n][1,n]=\lim_{x_{n+1}^2\rightarrow\infty} \frac{(x_{n+1}^2)^{\d[n+1]}}{\Gamma(\d[n+1])} \ \gcfta[1,n+1][1,n+1]
\end{equation}

The value of the conformal correlator in Eq.\eqref{flatfromsoft} is related via inversion to its value in the circumcentric configuration in Eq. \eqref{circconfig}, and hence \eqref{circconfig} implies \eqref{flatfromsoft}. This also explains the factor of $(x_{n+1}^2)^{\d[n+1]}$ needed in obtaining a finite value for the correlator as the point $x_{n+1}$ is taken to infinity. 

The relation between contact diagrams in \eqref{flatfromsoft} immediately implies a similar equality, presented in \eqref{softarbitfeyn} between every Feynman diagram contribution with $n$ external fields and $V$ vertices in the massless field theory, and its counterpart diagram contributing to a conformal correlator with  $(n+V)$ external fields and $V$ vertices. This is because at every vertex in any Feynman diagram, we can apply the relation \eqref{flatfromsoft} independently, thus proving our general result that all massless correlators can be expressed in terms of their conformal counterparts. 

We now provide a summary of the rest of the paper. 

In \S \ref{sec:contact}, we start with a general treatment of contact diagrams. We provide the first few examples of the equality \eqref{dequality} between $\dflat$ and $\dcft$ functions for three and four point contact massless correlators, as well as the soft limits and factorization for these cases. Next, we provide a proof of \eqref{circconfig} for all contact contributions i.e. for arbitrary number of fields and scaling dimensions $\d[i]$. This implies that the relation \eqref{flatfromsoft} also holds for all contact diagrams. 

In \S \ref{exch}, we provide a general treatment of all diagrams with one exchanged field, in other words, all diagrams with two vertices. We use this representation to prove \eqref{soft1exch} which is the counterpart of \eqref{flatfromsoft} for all diagrams with two vertices.

In \S \ref{genfeyn}, we provide a proof of the correspondence between all massless and conformal Feynman diagrams. To achieve this, we first describe a simple cutting rule that expresses any Feynman diagram as spacetime integrals over products of contact diagrams.

In \S \ref{Exact}, we use the proved correspondence to predict exact answers for the massless correlators. We start with the 5-point contact diagram and go on to provide examples for three and four point correlators including several loops. 

In \S \ref{massive}, we show that massive correlators can be expressed as infinite sums of conformal correlators. We conclude with a discussion of the results and several interesting questions for future work in \S \ref{disc}. The appendices contain examples of exact evaluations of contributions to conformal correlators that are used at several places in the main text. 

\section{Contact contributions}\label{sec:contact}
In this section, we begin with a general treatment of the contact contribution to the $n-$point correlator of massless fields. We use this to find explicit answers for the 3-point and 4-point correlator in \S \ref{firstexs}. We compare these answers with the 4-point and 5-point conformal contact contributions respectively, and find the first couple of examples of the equality claimed in \eqref{dequality} between the respective $\dflat$ and $\dcft$ functions. Next, in \S \ref{softfact}, we provide the details of the soft limit and factorization properties of the conformal correlators for these examples. In \S \ref{contactequality}, we provide the proof of \eqref{circconfig} for all contact correlators. 

Contact contributions to the $n-$point function coming from an interaction term of the form $\lambda \, \phi^n$ can be written as an integral over a bulk spacetime point $y$ as
\begin{equation} \label{ncontact}
	{\cal G}^{\flat}_{\d[1],\ldots,\d[n]}(x_1,\ldots,x_n) = \mathcal{A}\int \frac{d^d y}{\pi^{d/2}} \prod_{i=1}^{n} G_{\d[i]
}(x_i-y)
\end{equation}
where the propagator for a field with scaling dimension $\Delta_{i}$ is
\begin{equation} \label{prop}
	G_{\d[i]}(x_i-y)=\frac{\Gamma(\d[i])}{((x_i-y)^2)^{\d[i]}} 
\end{equation}
and $\mathcal{A}=\frac{i \lambda}{4^4 \pi^{3d/2}}$. For convenience, we shall drop this numerical factor $\mathcal{A}$ henceforth. We shall also drop the corresponding numerical factors in diagrams with multiple vertices that are treated in the rest of this article. We will work in Euclidean spacetime for convenience and all the results can be analytically continued to Lorentzian spacetime by the usual $i \epsilon$ prescription.

With the propagator coming from the usual kinetic term, the scaling dimensions are all $\Delta_i=\dm2$ when working around the free field fixed point. However, for now we will work with general scaling dimensions for simplicity. The resulting expressions are more general and can be used, for example, with scaling dimensions corresponding to renormalized propagators, or when working in dimensional regularization so that the spacetime dimensions are taken away from their integral values etc. We also find that the answers with arbitrary scaling dimensions $\d[i]$ are useful in computing the correlation functions of massive fields in \S \ref{massive}. 

We now turn to the evaluation of the $n$-point contact correlator in \eqref{ncontact}, carried out in \cite{Pibv} but included here for completeness. Integrals corresponding to the $n-$point contact diagram for the special case of conformally invariant interactions were computed in \cite{Symanzik:1972wj}. We will use a similar method, and highlight differences with the conformally invariant case later. Using Schwinger parametrization, \eqref{ncontact} takes the form
\begin{equation}
	\begin{split}
		\gflata[1,n][1,n]  &=  {1\over  \pi^{d/2} }\prod_{i=1}^n \left(\int_{0}^{\infty} dt_i t_i^{\Delta_i-1} \right) \int d^d y \ e^{-\sum_{i=1}^{n} t_i (x_i-y)^2}
	\end{split}
\end{equation}

Performing the Gaussian integral over the bulk point $y$ gives
\begin{equation}
	\begin{split}
		\gflata[1,n][1,n]&= \prod_{i=1}^n \int_{0}^{\infty} dt_i t_i^{\Delta_i-1} {1 \over T^{d/2}} e^{-{1\over T} \left(T \sum\limits_{i=1}^{n} t_i x_i^2-\left(\sum\limits_{i=1}^{n} t_i x_i\right)^2\right)},
	\end{split}
\end{equation}
where we have defined $T\equiv\sum\limits_{i=1}^n {t_i} $. 

Lorentz invariance dictates that the correlator should depend on the coordinates only as functions of $(x_i-x_j)^2$. Indeed, we see that
\begin{equation}
	\begin{split}
		T\sum_{i=1}^{n} t_i x_i^2-\left(\sum_{i=1}^{n}t_i x_i\right)^2 &=  \sum_{i=1}^{n}  t_i (T-t_i) x_i^2-\sum_{\substack{i,j=1 \\ i \neq j}}^{n} (t_i t_j x_i \cdot x_j) \\ 
		&=  \sum_{\substack{i,j=1 \\ i \neq j}}^{n}  \left( t_i t_j x_i^2-t_i t_j x_i \cdot x_j \right)=\sum_{\substack{i,j=1 \\ i \le j}}^{n} t_i t_j (x_i-x_j)^2
	\end{split}
\end{equation}
Defining $s_{ij}\equiv(x_i-x_j)^2$, we get
\begin{equation} \label{eq:schpara}
	\begin{split}
		\gflata[1,n][1,n]&= \prod_{i=1}^n \int_{0}^{\infty} dt_i t_i^{\Delta_i-1}{1 \over T^{d/2}} \exp(-{1\over T} \sum\limits_{\substack{i,j=1 \\ i \le j}}^{n} t_i t_j s_{ij}).
	\end{split}
\end{equation}
Henceforth, we will stop displaying the limits and the $i\leq j$ condition on the sum $\sum t_i t_j s_{ij}$. We now introduce unity in the form of $$ 1= \int_0^{\infty} ds \delta(s-\sum\limits_{i=1}^n \lambda_i t_i) $$ for arbitrary $\lambda_i \geq 0$ to find 

\begin{equation} 
	\begin{split}
		\gflata[1,n][1,n]&= \prod_{i=1}^n \int_{0}^{\infty} dt_i t_i^{\Delta_i-1}{1 \over T^{d/2}} e^{-{1\over T} \sum t_i t_j s_{ij}} \int_0^{\infty} ds \delta(s-\sum\limits_{i=1}^n \lambda_i t_i)
	\end{split}
\end{equation}
We substitute $t_i=s \alpha_i$ and use the notation $\D[n] \equiv \sum\limits_{i=1}^n \Delta_i$ to arrive at
\begin{equation} \label{eq:lamexp}
	\begin{split}
	\gflata[1,n][1,n]&=\prod_{i=1}^n \int_{0}^{\infty} d\alpha_i \alpha_i^{\Delta_i-1}{1 \over (\sum\limits_{i=1}^n \alpha_i)^{d/2}} \delta(1-\sum\limits_{i=1}^n \lambda_i \alpha_i)\\
		&\times \int_0^{\infty} ds  s^{\D[n]-1-d/2} \exp(-{s\over \sum\limits_{i=1}^n \alpha_i} \sum \alpha_i \alpha_j s_{ij})
	\end{split}
\end{equation}
We perform the integral over $s$ and obtain the $n$-point massless correlator in the form
\begin{equation} \label{schpolyn}
	\begin{split}
		\gflata[1,n][1,n]&=  \Gamma\left(\D[n]-{d\over2}\right)\prod_{i=1}^n \int_{0}^{\infty} d\alpha_i \alpha_i^{\Delta_i-1} \delta(1-\sum\limits_{i=1}^n \lambda_i \alpha_i)  {(\sum\limits_{i=1}^n \alpha_i)^{\D[n]-d} \over \left(\sum \alpha_i \alpha_j s_{ij}\right)^{\D[n]-d/2}}
	\end{split}
\end{equation}

If, instead, we are interested in the conformally invariant contact contribution, then the calculation proceeds exactly as above with the substitution $\D[n]=d$, so that the factor $\sum\limits_{i=1}^n \alpha_i^{\D[n]-d}$ in the numerator of \eqref{schpolyn} disappears, and we get 

\begin{equation} \label{schcftn}
	\begin{split}
		\gcfta[1,n][1,n]&=  \Gamma\left({d\over2}\right)\prod_{i=1}^n \int_{0}^{\infty} d\alpha_i \alpha_i^{\Delta_i-1} \delta(1-\sum\limits_{i=1}^n \lambda_i \alpha_i)  {1 \over \left(\sum \alpha_i \alpha_j s_{ij}\right)^{d/2}}
	\end{split}
\end{equation}
In this case, Symanzik \cite{Symanzik:1972wj} made the clever observation that, if one deformed the original expression \eqref{eq:schpara} by having $T=\sum \theta_i t_i$, with arbitrary $\theta_i \geq 0$, then similar manipulations as above land us again at exactly the expression in \eqref{schcftn}. This means that in \eqref{eq:schpara}, if one replaces $T=\sum t_i$ by $\sum \theta_i t_i$, the value of the integral remains the same! Thus, one is free to choose judicious values of the  $\theta_i$ to simplify the integrals. The expression in \eqref{schcftn} enjoys a similar freedom in the choice of values of $\lambda_i$. In Appendix \ref{confcontact}, we use this freedom to perform the integrals over $\alpha_i$ for up to six external fields. We also comment on how we organize this computation which makes the generalization to any number of external fields immediate.

As we noted in the introduction, the massless field correlators can be expressed in terms of certain building blocks that we denote by $\dflat$ functions. These are related via \eqref{dequality} to the $\dcft$ functions that build the conformal correlation functions. We shall now explicitly demonstrate this relation for the case of three and four external insertions. 
\subsection{First examples: Three and four point contact correlators} \label{firstexs}
\paragraph{3-point correlator:\\}

Massless correlators can be written in the form displayed in \eqref{gflat} as a prefactor times a function of ratios of distances. With three points $\{x_1,x_2,x_3\}$, two such ratios can be formed, for example

\begin{equation}
	r_1=\frac{s_{12}}{s_{23}}, \quad r_2=\frac{s_{13}}{s_{23}}	
\end{equation}

The contact contribution to the flat space 3-point massless correlator was found in \cite{Pibv} (in Appendix H) to be \footnote{We note that integrals of the kind \eqref{dflat3},\eqref{dcft4} were also arrived at in \cite{Boos:1990rg,Davydychev:1992mt}, albeit in slightly different contexts. They also derived the representation in terms of Appell F4 functions given in \eqref{dflat3f4}, as well as the simplification  \eqref{dflat3exact} in terms of dilogarithms in the special case. The representation in \eqref{dflat34d} was also found in \cite{Usyukina:1975yg}, in a related context. See also \cite{Ferrara:1974nf} for related work.}
\begin{equation} \label{gflat3}
	\begin{split}
		{\cal G}^{\flat}_{\Delta_1, \Delta_2,\Delta_3}(x_1, x_2, x_3)=& {1 \over \Gamma(d-\D[3]) }  {1 \over s_{23}^{\D[3]-d/2}} \, \dflat_{\Delta_1, \Delta_2,\Delta_3}(r_1,r_2), \\
	\end{split}
\end{equation}
where we have introduced the function $\dflat_{\Delta_1, \Delta_2,\Delta_3}(r_1,r_2)$ given by
\begin{equation} \label{dflat3}
	\begin{split}
		\dflat_{\Delta_1, \Delta_2,\Delta_3}(r_1,r_2)
		= \int  {da \over {2\pi i}} \int {db \over {2\pi i}} & \,r_1^a \, r_2^b \,\Gamma(-a) \Gamma(-b) \Gamma(\Delta_1+a+b) \Gamma(\sum_{i=1}^{3}\Delta_i-{d \over2}+a+b)  \\
		&\times \Gamma({d \over2}-\Delta_1 -\Delta_3-a) \Gamma({d \over2}-\Delta_1 -\Delta_2-b)
	\end{split}
\end{equation} 
If the conformal constraint \eqref{confconstraint} is satisfied, then the expression simplifies to the form of the familiar 3-point CFT correlator, \footnote{We find this answer in a couple of different ways in Appendix \ref{confcontact}.}
\begin{equation} 
	\begin{split}
		\gcft_{\Delta_1,\Delta_2,\Delta_3}(x_1, x_2 , x_3)&= \frac{\Gamma(\Delta_{12,3}) \Gamma(\Delta_{23,1}) \Gamma(\Delta_{13,2})}{\left(s_{12}\right)^{\half\Delta_{12,3}}(s_{23})^{\half\Delta_{23,1}}(s_{13})^{\half\Delta_{13,2}}} 
	\end{split}
\end{equation}
where $\Delta_{ij,k}\equiv\Delta_i+\Delta_j-\Delta_k$.

On the other hand, the contact contribution to the conformal 4-point correlator as a function of the cross ratios
\begin{equation} \label{cr4} s_1=u=\frac{s_{12}s_{34}}{s_{23}s_{14}}, \qquad  \qquad s_2=v=\frac{s_{24}s_{13}}{s_{23}s_{14}}
\end{equation}
can be computed, as shown in Appendix \ref{confcontact}, to be
\begin{equation} \label{gcft4}
	\begin{split}
		\gcfta[1,4][1,4]	&= \left(\frac{s_{23}}{s_{34}s_{24}}\right)^{\Delta_4-d/2} {1\over {s_{34}^{\Delta_3}s_{24}^{\Delta_2}s_{14}^{\Delta_1}}}  \dcft_{\Delta_1\Delta_2\Delta_3\Delta_4} (u,v)
	\end{split}
\end{equation}

with 
\begin{equation} \label{dcft4}
	\begin{split}
		\mathcal{D}^{cft}_{\Delta_1\Delta_2\Delta_3\Delta_4}(u,v)
		&=\int {ds \over 2\pi i} \int {dt \over 2\pi i} \Big\{ u^s v^t \, \Gamma(-s)\Gamma(-t)\,\Gamma({d\over2}+s+t-\Delta_4)\\
		&\times\Gamma(\Delta_{34}-{d\over2}-s)\,\Gamma(\Delta_1+s+t)\,\Gamma(\Delta_{24}-{d\over2}-t) \Big\},
	\end{split}
\end{equation}
where we have defined $\d[ij]=\d[i]+\d[j]$. This result \eqref{gcft4} was also computed in \cite{Dolan:2001}, presented as Eq. (B.8) there. 

We now have the ingredients in place to identify the first example of equality of the $\dflat$ and $\dcft$ functions. Replacing $\Delta_4= d-\sum_{i=1}^{3}\Delta_i$ turns \eqref{dcft4} into \eqref{dflat3} after the identifications $u \rightarrow r_1$ and $v \rightarrow r_2$, i.e.
\begin{equation}\label{d3equality}
		\dflat_{\d[1],\d[2],\d[3]}(r_1,r_2)=\mathcal{D}^{cft}_{\Delta_1,\Delta_2,\Delta_3,d-\D[3]}(u,v)|_{u=r_1,v=r_2}
\end{equation}
This proves the claimed equality \eqref{dequality} for $n=3$.  

We take a small detour to present the closed form expression for the correlator. The advantage with the Mellin-Barnes type of representation such as in \eqref{dflat3},\eqref{dcft4} is that performing the integrals leads to a series representation for the correlator. 
As presented in Appendix H of \cite{Pibv}, the integrals can be performed to find the 3-point $\dflat$ function in the form
\begin{equation} \label{dflat3f4}
	\begin{split}
		\dflat_{\d[1],\d[2],\d[3]}(r_1,r_2)=  
		 & \Gamma(\Delta_1) \Gamma(\D[3]-d/2) \Gamma(d/2-\Delta_{13}) \Gamma(d/2-\Delta_{12})\\&\times F_4(\Delta_1,\D[3]-d/2,1-d/2-\Delta_{13},1-d/2-\Delta_{12};r_1,r_2) \\
		+&r_2^{d/2-\Delta_{12}} \Gamma(\Delta_3) \Gamma(d/2-\Delta_2)  \Gamma(d/2-\Delta_{13})\Gamma(\Delta_{12}-d/2) \\
		&\times F_4(\Delta_3,d/2-\Delta_2,1-d/2-\Delta_{13},1+d/2+\Delta_{12};r_1,r_2)\\
		+& r_1^{d/2-\Delta_{13}} \,\Gamma(\Delta_2) \Gamma(d/2-\Delta_3) \Gamma(d/2-\Delta_{12}) \Gamma(\Delta_{13}-d/2)\\& \times F_4(\Delta_2,d/2-\Delta_3,1+d/2+\Delta_{13},1-d/2-\Delta_{12};r_1,r_2)\\
		+&r_1^{d/2-\Delta_{13}} r_2^{d/2-\Delta_{12}}\, \Gamma(d-\D[3])\Gamma(d/2-\Delta_1)\Gamma(\Delta_{12}-d/2)\Gamma(\Delta_{13}-d/2) \\ \times &F_4(d-\D[3],d/2-\Delta_1,1+d/2+\Delta_{13},1+d/2+\Delta_{12};r_1,r_2), 
	\end{split}
\end{equation}
in terms of the Appell $F_4$ function of the simple ratios  $r_1=\frac{s_{12}}{s_{13}}$ and $r_2=\frac{s_{14}}{s_{13}}$ \footnote{ The Appell F4 function is defined by
\begin{equation}
	F_4(a,b,c,d;x,y)=\sum_{m,n=0}^{\infty} \frac{(a)_{m+n}(b)_{m+n}}{c_{(m)} d_{(n)}} \frac{x^m y^n}{m! n!}
\end{equation}.}.
As a check of our result \eqref{d3equality}, we find that the $\dflat$ function in \eqref{dflat3f4} is the same as the 4-point $\dcft$ function obtained by Dolan and Osborn in Eq. (B.10) of \cite{Dolan:2001}. 


As shown in \cite{Dolan:2001}, the expression in  \eqref{dcft4} simplifies when each $\Delta_i=1$, realized when working around the free field fixed point in $d=4$. We now use this result to find a simpler form for the 3-point massless correlator in this case. With $\Delta_i=1$, the expression in \eqref{dflat3} takes the form
\begin{equation} \label{dflat34d}
	\begin{split}
		\dflat_{1,1,1}(r_1,r_2)
		= \int  {da \over {2\pi i}} \int {db \over {2\pi i}} & \,r_1^a \, r_2^b \,\Gamma^2(-a) \Gamma^2(-b) \Gamma^2(1+a+b) 
	\end{split}
\end{equation} 
Using Eq. (C.15) of  \cite{Dolan:2001}, we find that the 3-point contact correlator generated by the Lagrangian $$\half \partial_{\mu} \phi \partial^{\mu}\phi +\lambda \phi^3$$ in four spacetime dimensions is given by 

\begin{equation} \label{dflat3exact}
	\begin{split}
		\gflat_{1,1,1}(x_1,x_2,x_3)=& {i \lambda\over {4^4 \pi^{6} s_{23}}} \frac{1}{(w-\bar{w})} \left(2 \text{Li}_2(w)-2\text{Li}_2(\bar{w})+\ln(w \bar{w}) \ln \left(\frac{1-w}{1-\bar{w}}\right) \right)\\
	\end{split}
\end{equation}
with the definitions $r_1=\frac{s_{12}}{s_{23}}=w \bar{w}$ and $r_2=\frac{s_{13}}{s_{23}}=(1-w)(1-\bar{w})$, and the reintroduction of the overall numerical factor just this once.

\paragraph{4-point correlator:\\} 

We now move to the case of four points. Now, the flat space answer is a function of five simple ratios which we choose to be

\begin{equation}
	r_1=\frac{s_{34}}{s_{14}}, r_2=\frac{s_{23}}{s_{14}}, r_3=\frac{s_{13}}{s_{14}},r_4=\frac{s_{24}}{s_{14}},r_5=\frac{s_{12}}{s_{14}}
\end{equation}
Contact contribution to the flat space 4-point function is 
\begin{equation} \label{gflat4}
	\begin{split}
		{\cal G}^{\flat}_{\Delta_1, \Delta_2,\Delta_3,\Delta_4}(x_1, \ldots,x_4)=& {1 \over \Gamma(d-\D[4]) }  {1 \over s_{14}^{\D[4]-d/2}} \, \dflat_{\Delta_1, \Delta_2,\Delta_3,\Delta_4}(r_1,\ldots,r_5), \\
	\end{split}
\end{equation}
with $\dflat_{\Delta_1, \Delta_2,\Delta_3,\Delta_4}(r_1,\ldots,r_5)$ given by
\begin{equation} \label{dflat4}
	\begin{split}
		\dflat_{\Delta_1, \Delta_2,\Delta_3,\Delta_4}(r_1,\ldots,r_5)
		= &\prod_{i=1}^{5}\int  {da_i \over {2\pi i}}   \,r_i^{a_i}  \,\Gamma(-a_i)  \Gamma(\Delta_2+a_{245})\Gamma(\Delta_3+a_{123})   \\
		&\times \Gamma({d \over2}-\Delta_{234}-a_{124}) \Gamma({d \over2}-\Delta_{123}-a_{235} )\Gamma(\sum_{i=1}^{4}\Delta_i-{d \over2}+\A[5])
	\end{split}
\end{equation} 
where we have defined the symbols $a_{ijk}=a_{i}+a_{j}+a_{k},\Delta_{ijk}=\Delta_{i}+\Delta_{j}+\Delta_{k},\A[n]=\sum_{i=1}^{n}a_i$.

On the other hand, the contact contribution to the conformal 5-point correlator can be expressed in terms of five cross ratios which we take to be 
\begin{equation} \label{cr5}
	s_1=\frac{s_{14}s_{25}}{s_{12}s_{45}}, s_2=\frac{s_{24}s_{15}}{s_{12}s_{45}},s_3=\frac{s_{15}s_{25}s_{34}}{s_{12}s_{35}s_{45}},s_4=\frac{s_{13}s_{25}}{s_{12}s_{35}},s_5=\frac{s_{23}s_{15}}{s_{35}s_{12}}
\end{equation}
As shown in Appendix \ref{confcontact}, the 5-point function takes the form
\begin{equation} \label{gcft5}
\gcfta[1,5][1,5]=\left(\frac{s_{12}}{s_{15}s_{25}}\right)^{\Delta_5-d/2}  \frac{1}{\prod_{i=1}^{4}s_{i5}^{\Delta_i}}\dcft_{\Delta_1,\ldots,\Delta_5}(s_1,\ldots,s_5)
\end{equation}
with 
\begin{equation} \label{dcft5}
	\begin{split}
		\dcft_{\Delta_1,\ldots,\Delta_5}(s_1,\ldots,s_5)
		= &\prod_{i=1}^{5}\int  {da_i \over {2\pi i}}   \,s_i^{a_i}  \,\Gamma(-a_i)  \Gamma(\Delta_3+a_{345})\Gamma(\Delta_4+a_{123})   \\
		&\times \Gamma\left(\Delta_{15}-a_{235}-{d \over2}\right)\Gamma\left(\Delta_{25}-a_{134}-{d \over2}\right) \Gamma\left(\A[5]+{d \over2}-\Delta_5\right)
	\end{split}	
\end{equation}
We now demonstrate a second example of the equality between $\dflat$ and $\dcft$ functions presented in \eqref{dequality}.
Substituting $\Delta_5= d-\sum_{i=1}^{4
}\Delta_i$ and  performing the replacement $s_i \rightarrow r_i$ in \eqref{dcft5}, we find agreement with \eqref{dflat4} 
  so that
\begin{equation}\label{d4equality}
\dflat_{\d[1],\ldots,\d[4]}(r_1,\ldots,r_5)=\mathcal{D}^{cft}_{\Delta_1,\ldots,\Delta_4,d-\D[4]}(s_1,\ldots,s_5)|_{s_1=r_1,\ldots,s_5=r_5}
\end{equation}

\subsection{Soft limit and factorization} \label{softfact}
We will demonstrate the proposed position space soft limit for the three and four point conformally invariant functions. 

Let us recall the expressions for the 3-point massless correlator in \eqref{gflat3},
\begin{equation}\label{gflat31}
	\begin{split}
		{\cal G}^{\flat}_{\Delta_1, \Delta_2,\Delta_3}(x_1, x_2, x_3)=&  {1 \over \Gamma(d-\D[3]) } {1 \over s_{23}^{\D[3]-d/2}} \, \dflat_{\Delta_1, \Delta_2,\Delta_3}(r_1,r_2), \\
	\end{split}
\end{equation}
and for the 4-point conformally invariant correlator in \eqref{gcft4},

\begin{equation}\label{gcft41}
	\begin{split}
	\gcfta[1,4][1,4]	&= \left(\frac{s_{23}}{s_{34}s_{24}}\right)^{\Delta_4-d/2} {1\over {s_{34}^{\Delta_3}s_{24}^{\Delta_2}s_{14}^{\Delta_1}}}  \dcft_{\Delta_1\Delta_2\Delta_3\Delta_4} (u,v)
	\end{split}
\end{equation}

In the circumcentric configuration with $x_4$ placed at the centre of the unit sphere circumscribing the points $x_1$, $x_2$, and $x_3$, i.e. with $s_{14}=s_{24}=s_{34}=1$, we find that
\begin{equation}
	 \begin{split}
	 	u=\frac{s_{12}s_{34}}{s_{23}s_{14}}=\frac{s_{12}}{s_{23}}=r_1, \qquad v=\frac{s_{24}s_{13}}{s_{23}s_{14}}=\frac{s_{13}}{s_{23}}=r_2
	 \end{split}
\end{equation}
The identification of the cross ratios with the simple ratios is realized in this configuration and so, as observed in  \eqref{d3equality}, leads to the equality of the 3-point $\dflat$ function in \eqref{gflat31}with the 4 point $\dcft$ function in \eqref{gcft41} using $\d[4]=d-\D[3]$. In addition, after this substitution, the prefactor in \eqref{gcft41} turns precisely into the prefactor in \eqref{gflat31}, thus demonstrating an example of the equality advocated in \eqref{circconfig} between massless and conformal correlators.

We are now in a position to derive the soft limit of the 4-point conformal correlator as the value of the correlator obtained after inverting the unit circumcentric configuration. We find that our position space soft limit of the 4-point conformal correlation function is
\begin{equation}\label{soft3}
\lim_{x_{4}^2\rightarrow\infty} \gcft_{\d[1],\d[2],\d[3],d-\Delta^{(3)}}(x_1,x_2,x_3,x_{4})= {{\Gamma(d-\D[3])}\over {s_{34}^{\Delta_{34}-d/2}s_{24}^{\Delta_{24}-d/2}s_{14}^{\Delta_1}}} \gflata[1,3][1,3]
\end{equation}
where the left hand side is defined via inversion of the circumcentric configuration defined below \eqref{gcft41}. The leading part of the prefactor above is then $(1/x_4^2)^{2\d[4]+\D[3]-d}=(1/x_4^2)^{\d[4]}$.
This means that we obtain the 3-point massless correlator as a soft limit of the 4-point conformal correlator via
\begin{equation}\label{flatfromsoft3}
	\gflata[1,3][1,3]=\lim_{x_{4}^2\rightarrow\infty} \frac{(x_{4}^2)^{\d[4]}} {\Gamma(d-\D[3])} \gcft_{\d[1],\d[2],\d[3],d-\Delta^{(3)}}(x_1,x_2,x_3,x_{4})
\end{equation}

We will now move on and consider the soft limit of the $5$-point conformal correlator. First, we recollect, from \eqref{gcft5}, the expression for the $5$-point conformal correlator,
\begin{equation} \label{gcft51}
	{\cal G}^{cft}_{\Delta_1,\ldots,\Delta_5}(x_1,\ldots,x_5)=\left(\frac{s_{12}}{s_{15}s_{25}}\right)^{\Delta_5-d/2}  \frac{1}{\prod_{i=1}^{4}s_{i5}^{\Delta_i}}\dcft_{\Delta_1,\ldots,\Delta_5}(s_1,\ldots,s_5)
\end{equation}
which depends on the cross ratios
\begin{equation} \label{cr5}
	s_1=\frac{s_{14}s_{25}}{s_{12}s_{45}}, s_2=\frac{s_{24}s_{15}}{s_{12}s_{45}},s_3=\frac{s_{15}s_{25}s_{34}}{s_{12}s_{35}s_{45}},s_4=\frac{s_{13}s_{25}}{s_{12}s_{35}},s_5=\frac{s_{23}s_{15}}{s_{35}s_{12}}
\end{equation}
We also recollect the expression for the $4$-point massless correlator in \eqref{gflat4} as
\begin{equation} \label{gflat41}
	\begin{split}
	\gflata[1,4][1,4]=&  {1 \over s_{14}^{\D[4]-d/2}} \, \dflat_{\Delta_1, \ldots,\Delta_4}(r_1,\ldots,r_5), \\
	\end{split}
\end{equation}
which is a function of the simple distance ratios,
\begin{equation}\label{sr4}
	r_1=\frac{s_{34}}{s_{14}}, r_2=\frac{s_{23}}{s_{14}}, r_3=\frac{s_{13}}{s_{14}},r_4=\frac{s_{24}}{s_{14}},r_5=\frac{s_{12}}{s_{14}}
\end{equation}
As is evident, in the limit of $x_5^2 \to 	\infty$, the cross ratios in $\eqref{cr5}$ turn into the simple ratios in $\eqref{sr4}$, and as noted in \eqref{d4equality}, the $\dcft$ function in \eqref{gcft51} matches with the $\dflat$ function in \eqref{gflat41}.  In the soft limit, the $5$-point conformal correlator exhibits the factorization,
\begin{equation}\label{soft4}
	\lim_{x_{5}^2\to\infty}\gcfta[1,5][1,5]= {{\Gamma(d-\D[4])} \over {s_{15}^{\Delta_{15}-d/2}s_{25}^{\Delta_{25}-d/2}s_{35}^{\Delta_3}s_{45}^{\Delta_4}}} \, \gflata[1,4][1,4]
\end{equation}
The expression for the $4$-point massless correlator can be easily obtained as
\begin{equation}\label{flatfromsoft4}
	\gflata[1,4][1,4]=\lim_{x_{5}^2\rightarrow\infty} \frac{(x_{5}^2)^{\d[5]}}{{\Gamma(d-\D[4])} }  \gcft_{\d[1],\d[2],\d[3],\d[4],d-\Delta^{(4)}}(x_1,\ldots,x_5)
\end{equation}
It is also evident that in the circumcentric configuration with $x_5$ at the centre of the sphere on which $x_1,\ldots,x_4$ lie, the $5$-point conformal correlator in \eqref{gcft51} exactly takes the form of the $4$-point flat space massless correlator in \eqref{gflat41}. After having described the first couple of examples in some detail, in the next subsection, we present a proof of the equality between massless and conformal correlators for all contact diagrams.

\subsection{Proof of equality of contact correlation functions} \label{contactequality}

We will now provide a proof of \eqref{circconfig} for the equality of contact contributions of the $n$-point massless correlator with the $(n+1)$-point conformal correlator. To do this, we use the representation of the respective correlation functions derived in \eqref{schpolyn} and \eqref{schcftn}. In particular, using  \eqref{schcftn}, we have
\begin{equation} 
	\begin{split}
		\gcfta[1,n+1][1,n+1]&=\Gamma(d/2)\left(\prod_{i=1}^{n+1} \int_{0}^{\infty} d\alpha_i \alpha_i^{\Delta_i-1} \right)\delta(1-\sum\limits_{j=1}^{n+1} \lambda_j \alpha_j)  {1 \over \left(\sum\limits_{\substack{i,j=1 \\ i \le j}}^{n+1}\alpha_i \alpha_j s_{ij}\right)^{d/2}}
	\end{split}
\end{equation}
We choose $\lambda_i=\delta_{i,n+1}$ i.e $\lambda_i=0$ for all $i\neq n+1$ and $\lambda_{n+1}=1$. The integral over $\alpha_{n+1}$ is then trivial and we get
\begin{equation} 
	\begin{split}
			\gcfta[1,n+1][1,n+1]&=\Gamma(d/2) \prod_{i=1}^{n} \int_{0}^{\infty} d\alpha_i \alpha_i^{\Delta_i-1}   {1 \over \left(\sum\limits_{i=1}^{n}\alpha_i s_{i (n+1)}+\sum\limits_{\substack{ i,j \neq n+1}} \alpha_i \alpha_j s_{ij}\right)^{d/2}}
	\end{split}
\end{equation}

We now introduce $$ 1= \int_0^{\infty} ds \delta(s-\sum\limits_{i=1}^{n} \lambda_i \alpha_i) $$ for arbitrary $\lambda_i \geq 0$, and then scale $\alpha_i \rightarrow s \alpha_i$ to find 

\begin{equation}
	\begin{split}
		\gcfta[1,n+1][1,n+1]=&\Gamma(d/2)\prod_{i=1}^{n} \int_{0}^{\infty} d\alpha_i \frac{\alpha_i^{\Delta_i-1}}{(\sum\limits_{i=1}^{n}\alpha_i s_{i (n+1)})^{d/2}} \delta(1-\sum\limits_{i=1}^{n} \lambda_i \alpha_i) \\
		& \times \int_0^{\infty} ds s^{\D[n]-d/2-1}   \left(1+s \frac{\sum\limits_{\substack{ i,j \neq n+1}} \alpha_i \alpha_j s_{ij}}{\sum\limits_{i=1}^{n}\alpha_i s_{i (n+1)}}\right)^{-d/2}
	\end{split}
\end{equation}
Performing the elementary integral over $s$ using \eqref{elemints},
\begin{equation} \label{cftnminusone}
	\begin{split}
		\gcfta[1,n+1][1,n+1]=&\Gamma(\D[n]-d/2) \Gamma(d-\D[n])\\
		&\times \prod_{i=1}^{n} \int_{0}^{\infty} d\alpha_i  \alpha_i^{\Delta_i-1}  \delta(1-\sum\limits_{i=1}^{n} \lambda_i \alpha_i)  \frac{(\sum\limits_{i=1}^{n}\alpha_i s_{i (n+1)})^{\D[n]-d}}{\left(\sum\limits_{\substack{ i,j \neq n+1}} \alpha_i \alpha_j s_{ij}\right)^{\D[n]-d/2}}
	\end{split}
\end{equation}
Now, we move to the circumcentric configuration with $x_{n+1}$ at the centre of a sphere of unit radius on which all the other insertions $x_1,\ldots,x_n$ lie, i.e. we go to the configuration with 

\begin{equation}
	s_{i(n+1)}=1 \qquad i=1,\ldots,n
\end{equation}

We then see that (up to the prefactor $\Gamma(d-\D[n])$ ), the right hand side precisely turns into \eqref{schpolyn} for the $n$-point massless correlator, so that we prove, for contact contributions 
\begin{equation} \label{gequalitycontact}
	\gflata[1,n][1,n]= \frac{1}{ \Gamma(d-\D[n])} \gcft_{\d[1],\ldots,\d[n],d-\Delta^{(n)}}(x_1,\ldots,x_n,0)|_{x_1^2=\ldots=x_n^2=1}
\end{equation}
 We can also provide a direct proof of \eqref{flatfromsoft} on similar lines as the one in this subsection. However, as noted in the introduction, \eqref{gequalitycontact} implies that the flat correlator can be obtained from the soft limit of the conformal correlator via \eqref{flatfromsoft}. In the next section, for variety, we provide a direct proof for the fact that massless 1-exchange correlators can be obtained as a soft limit of their conformal counterparts, i.e. the 1-exchange counterpart of \eqref{flatfromsoft}.

\section{1-Exchanges} \label{exch}
\begin{figure}
	\centering  \includegraphics[height=.2\textheight]{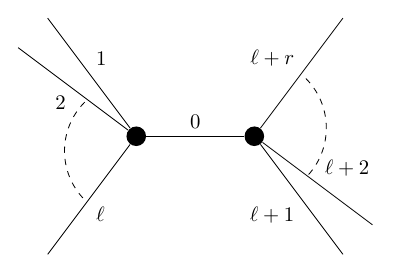}
		\caption{The general 1-exchange diagram, with $\ell$ and $r$ external fields at the left and right vertices respectively, for a total of $n=\ell+r$ external fields. The displayed numbers are the indices that label both the scaling dimensions and Schwinger parameters for each field.}
	\label{1exchlr}
\end{figure}

In this section, we consider diagrams with one exchanged field between two interaction vertices. Let's consider the most general such diagram shown in Figure \ref{1exchlr}. 
We denote by $\ell/r$ the number of fields interacting at the left/right vertex whose scaling dimensions are $\{\d[1],\ldots,\d[\ell]\}$ and $\{\d[\ell+1],\ldots,\d[\ell+r]\}$ respectively. We denote the scaling dimension of the exchanged field by $\d[0]$, and the total number of external fields by $n=\ell+r$. The contribution to the massless $n$-point correlator from such a diagram is given by

\begin{equation} \label{gflaten}
{\gflat}_{e,n} ={1 \over \pi^d}\int d^d y_1 \left(\prod_{i=1}^{\ell}\frac{\Gamma(\d[i])}{((x_i-y_1)^2)^{\Delta_i}}\right) \int d^d y_2 \left(\prod_{i=\ell+1}^{\ell+r}\frac{\Gamma(\d[i])}{((x_i-y_2)^2)^{\Delta_i}} \right) \frac{\Gamma(\d[0])}{((y_1-y_2)^2)^{\Delta_0}},   
\end{equation}
where for brevity, on the LHS, we have suppressed the dependence of the correlator on the scaling dimensions and the spacetime points, and as in the contact case, dropped an overall numerical factor proportional to square of the coupling constant.

Before evaluating this integral, we establish some convenient notation. We introduce Schwinger parameters $t_0$ for the internal propagator and $\{t_1,\ldots,t_{\ell}\}$ for all the other propagators involving $y_1$.  We denote the set of all indices associated with the fields interacting at the left vertex  $y_1$ by $L=\{0,1,\ldots,\ell\}$, and the ones at the right vertex $y_2$ by $R=\{0,\ell+1,\ldots,\ell+r\}$. We use the notation $T_L$ and $T_R$ for the sums of the Schwinger parameters corresponding to the sets $L$ and $R$ respectively. Similarly, we denote by $\D[L]$ and $\D[R]$ the sums of the scaling dimensions of fields corresponding to the sets  $L$ and $R$ respectively.  We also denote the sets of left and right indices omitting the one on the exchanged field by $L'=\{1,\ldots,\ell\}$ and $R'=\{\ell+1,\ldots,\ell+r\}$ respectively. 

The integral over $y_1$ in \eqref{gflaten} is similar to the one in the contact contribution in \S \ref{sec:contact}. Performing this Gaussian integral, we find 
\begin{equation} \label{lint}
{\gflat}_{e,n} ={1 \over \pi^{d/2}} \int d^d y_2 \left(\prod_{i=\ell+1}^{\ell+r}\frac{\Gamma(\d[i])}{((x_i-y_2)^2)^{\Delta_i}} \right)    \prod_{i=0}^{\ell} \int_{0}^{\infty} dt_i t_i^{\Delta_i-1}{1 \over T_{L}^{d/2}} \exp(-{1\over T_{L}} \sum\limits_{\substack{i,j=0 \\ i \le j}}^{\ell} t_i t_j s_{ij}),
\end{equation}
We make the following variable substitution for the Schwinger parameters in the set $L$,
\begin{equation} \label{schsub}
t'_i={t_i \over T_L^{1/2}} \qquad  i \in L
\end{equation}
It is easy to check that the sum of the new Schwinger parameters is
$T'_L=\sum_{i\in L} t'_i=T_L^{1/2}$,
and that the determinant of the Jacobian involved in the transformation to the new Schwinger parameters is $2 (T'_L)^{\ell+1}$. After the variable substitution then, we find
\begin{equation} 
{\gflat}_{e,n} ={1 \over \pi^{d/2}} \int d^d y_2 \left(\prod_{i=\ell+1}^{\ell+r}\frac{\Gamma(\d[i])}{((x_i-y_2)^2)^{\Delta_i}} \right)    \prod_{i=0}^{\ell} \int_{0}^{\infty} dt'_i (t'_i)^{\Delta_i-1}{1 \over (T'_{L})^{d-\Delta_L}} \exp(- \sum\limits_{\substack{ L }} t'_i t'_j s_{ij}),
\end{equation}
where we have introduced the shorthand notation $\sum\limits_{\substack{ L }}$ for the sum $\sum\limits_{\substack{i,j=0 \\ i \le j}}^{\ell}$. We now drop the tildes on the Schwinger parameters. 

We introduce Schwinger parameters $\{t_{\ell+1},\ldots,t_{n}\}$ for all the other propagators involving $y_2$, and perform the integral over $y_2$ to see that
\begin{equation} \label{lrint}
{\gflat}_{e,n} =   \prod_{i=0}^{n} \int_{0}^{\infty} dt_i t_i^{\Delta_i-1}{1 \over T_{L}^{d-\Delta_L}}e^{- \sum\limits_{ L' } t_i t_j s_{ij}}  {1 \over W^{d/2}}e^{- {1 \over W}\sum\limits_{L'+R' } w_i w_j s_{ij}}
\end{equation}
where we have introduced the variables 
\begin{equation}
w_i = \begin{cases}
t_0 t_i & i\in L' \\
t_i & i \in R'
\end{cases}
\end{equation}
and their sum by $W=\sum\limits_{i \in L'+R'} w_i$. We now make one final substitution for all the Schwinger parameters to the following set
\begin{equation}
a_i = \begin{cases}
{t_0 \over W^{1/2}} & i=0 \\
{t_i \over W^{1/2}} & i \in R' \\
t_i & i \in L'
\end{cases}
\end{equation}
We also define the sum $A=\sum\limits_{i\in L'} a_0 a_i +\sum\limits_{i\in R'} a_i$ and find that $A=W^{1/2}$. The determinant of the Jacobian needed to make the transformation to the $a_i$ parameters is given by $2 A^{r+1}$. Performing this transformation, we find that the integral representation of the exchange diagram (the counterpart of the representation of the contact diagram in \eqref{eq:schpara}) is
\begin{equation} \label{flat1e}
{\gflat}_{e,n} =   \prod_{i=0}^{n} \int_{0}^{\infty} da_i a_i^{\Delta_i-1}{1 \over (a_0 A+A_{L'})^{d-\Delta_L}}e^{- \sum\limits_{ L' } a_i a_j s_{ij}}  {1 \over A^{d-\Delta_R}}e^{- \sum\limits_{L'+R'} a'_i a'_j s_{ij}}
\end{equation}
where we have introduced
\begin{equation}\label{aprime}
a'_i = \begin{cases}
a_0 a_i & i\in L' \\
a_i & i \in R'
\end{cases}
\end{equation}
and $A_L'=\sum\limits_{i\in L'} a_i$.
\paragraph{Proof of equality for 1-exchange diagrams\\}

Here, we will show that the massless correlator in Fig.\ref{1exchlr} having two vertices, one exchanged field and an arbitrary number of external fields is the same as a conformal correlator with two extra fields, one at each vertex. We will show this by transforming the integral representation for the 1-exchange diagram of $n$ massless fields in \eqref{flat1e} into the conformally invariant 1-exchange diagram with $(n+2)$ fields.  

First, for convenience in what follows, we relabel all the external fields at the right vertex in \eqref{flat1e}. This relabelling is just an increase of each of their indices by one, so that we now have the set of indices of all fields at the right vertex to be $R=\{0,\ell+2,\ldots,n+1\}$, and the corresponding set with only external fields to be $R'=\{\ell+2,\ldots,n+1\}$. We now introduce $\d[\ell+1]=d-\Delta_L$ and $\d[n+2]=d-\Delta_R$. As the notation indicates, these will turn out to be the scaling dimensions of the extra fields. We multiply and divide \eqref{flat1e} by $ (x_{\ell+1}^2)^{\d[\ell+1]} (x_{n+2}^2)^{\d[n+2]}$, where $x_{\ell+1}$ and $x_{n+2}$ will turn out to be the locations of the extra fields. We then introduce two extra Schwinger parameters $a_{\ell+1}$ and $a_{n+2}$, which will end up being the Schwinger parameters corresponding to the two extra propagators in the conformally invariant 1-exchange diagram. We use these extra Schwinger parameters to exponentiate the two denominators in \eqref{flat1e} and obtain
\begin{equation}\label{flat1e1} 
\begin{split}
{\gflat}_{e,n} = {(x_{\ell+1}^2)^{\d[\ell+1]} (x_{n+2}^2)^{\d[n+2]}\over \Gamma(\d[\ell+1]) \Gamma(\d[n+2])}  & \prod_{i=0}^{n+2} \int_{0}^{\infty} da_i a_i^{\Delta_i-1}\int_0^{\infty}da_{\ell+1}\, a_{\ell+1}^{\d[\ell+1]-1} e^{-a_{\ell+1}(a_0 A+A_{L'}) x_{\ell+1}^2 } \\
\times & \int_0^{\infty}da_{n+2} \, a_{n+2}^{\d[n+2]-1} e^{-a_{n+2} A \, x_{n+2}^2 } e^{- \sum\limits_{ L' } a_i a_j s_{ij}}  e^{- \sum\limits_{L'+R'} a'_i a'_j s_{ij}}
\end{split}
\end{equation}
When compared to \eqref{flat1e}, the above expression has an extra exponential term given by
\begin{equation}\label{extraexp}
e^{-a_{\ell+1}(a_0 A+A_{L'}) x_{\ell+1}^2 -a_{n+2} A \, x_{n+2}^2 }
\end{equation}
We manipulate its exponent as follows
\begin{equation}
\begin{split}
 &-a_{\ell+1}(a_0 A+A_{L'}) x_{\ell+1}^2 -a_{n+2} A \, x_{n+2}^2\\
 =& -a_{\ell+1}\sum\limits_{i \in L'} a_i x_{\ell+1}^2 -a_{\ell+1}a_0 (a_0 \sum\limits_{i \in L'} a_i+\sum\limits_{i \in R'} a_i) x_{\ell+1}^2  -a_{n+2} (a_0 \sum\limits_{i \in L'} a_i+\sum\limits_{i \in R'} a_i)  \, x_{n+2}^2\\
 =& -a_{\ell+1}\sum\limits_{i \in L'} a_i x_{\ell+1}^2 -a'_{\ell+1} \sum\limits_{i \in L'} a'_i \, x_{\ell+1}^2-a'_{\ell+1} \sum\limits_{i \in R'} a_i x_{\ell+1}^2 - a_{n+2} \sum\limits_{i \in L'} a'_i x_{n+2}^2  -a_{n+2} \sum\limits_{i \in R'} a_i  x_{n+2}^2
\end{split}
\end{equation}
 where we have introduced $a'_{\ell+1}=a_0 a_{\ell+1}$, naturally continuing the definition in \eqref{aprime}. We notice that the exponential in \eqref{extraexp} can also be written as
 \begin{equation}\label{extraexp1}
 \begin{split}
 e^{-a_{\ell+1}(a_0 A+A_{L'}) x_{\ell+1}^2 -a_{n+2} A \, x_{n+2}^2 }= \lim\limits_{x_{\ell+1}^2,x_{n+2}^2\to \infty} & e^{-a_{\ell+1} \sum\limits_{i \in L'} a_i s_{i (\ell+1)} -a'_{\ell+1} \sum\limits_{i \in L'} a'_i s_{i (\ell+1)}} \\
 &e^{-a'_{\ell+1} \sum\limits_{i \in R'} a_i s_{i (\ell+1)} -a_{n+2} \sum\limits_{i \in L'} a'_i s_{i (n+2)}}e^{-a_{n+2} \sum\limits_{i \in R'} a_i s_{i (n+2)}}  
 \end{split}
 \end{equation}
  We now introduce $L'_+=\{1,\ldots,\ell+1\}$ and $R'_{+}=\{\ell+2,\ldots,n+2\}$, and see that the exponentials in \eqref{extraexp1} combine nicely with the other exponentials in \eqref{flat1e1} to produce
  \begin{equation}\label{flat1e2} 
  \begin{split}
  {\gflat}_{e,n} = \lim\limits_{x_{\ell+1}^2,x_{n+2}^2\to \infty}  {(x_{\ell+1}^2)^{\d[\ell+1]} (x_{n+2}^2)^{\d[n+2]}\over \Gamma(\d[\ell+1]) \Gamma(\d[n+2])} & \prod_{i=0}^{n+2} \int_{0}^{\infty} da_i a_i^{\Delta_i-1} e^{- \sum\limits_{ L'_{+} } a_i a_j s_{ij}}  e^{- \sum\limits_{L'_{+}+R'_{+}} a'_i a'_j s_{ij}}
  \end{split}
  \end{equation}
with the natural generalization of \eqref{aprime} to
\begin{equation}\label{aprimep}
	a'_i = \begin{cases}
		a_0 a_i & i\in L_{+}' \\
		a_i & i \in R_{+}'
	\end{cases}
\end{equation}
  We now observe that the integral above is the same as that for a conformally invariant 1-exchange diagram with the external fields at the left vertex having scaling dimensions $\{\d[1],\ldots,\d[\ell+1]\}$, the external fields at the right vertex having scaling dimensions  $\{\d[\ell+2],\ldots,\d[n+2]\}$, and the exchanged field having scaling dimension $\d[0]$. The answer for this conformally invariant 1-exchange diagram can be obtained by starting from the expression in \eqref{flat1e}, adding one extra field at each vertex, and putting in the conformality constraints at each vertex that set both denominators to unity. Hence, we have shown that
  \begin{equation} \label{soft1exch}
  \gflat_{e,n}=\lim\limits_{x_{\ell+1}^2,x_{n+2}^2\to \infty}  {(x_{\ell+1}^2)^{\d[\ell+1]} (x_{n+2}^2)^{\d[n+2]}\over \Gamma(\d[\ell+1]) \Gamma(\d[n+2])}  \gcft_{e,n+2}. 
  \end{equation}

 \section{General Feynman diagrams} \label{genfeyn}
In this section, we treat the most general Feynman diagram contribution to the massless and conformal correlators we have been studying. In \S \ref{cutting}, we derive a \emph{cutting rule} that expresses any general Feynman diagram as spacetime integrals over products of contact diagrams. In \S \ref{allfeyneq}, we present a proof for the equality between the contribution of a general Feynman diagram with $V$ vertices to the $n-$point massless correlator and its counterpart contribution with $V$ vertices to the $(n+V)-$point conformal correlator. 

\subsection{Cutting rule in position space} \label{cutting}
\begin{figure}
	\centering  \includegraphics[height=2.1cm]{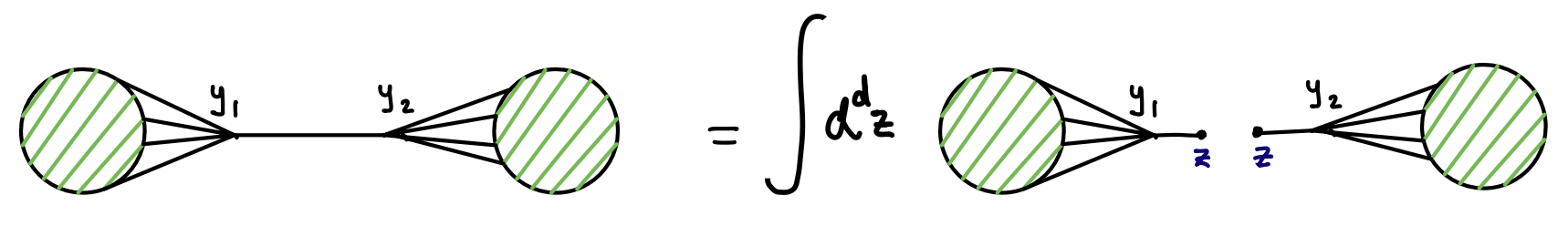}
	\caption{Depiction of the cutting rule}
	\label{cutting1}
\end{figure}

We now derive a cutting rule for Feynman diagrams of massless fields depicted in Fig.\ref{cutting1}. 
The figure on the left represents a generic Feynman diagram where we focus on an internal line connecting two vertices, one at $y_1$ and the other at $y_2$, and with scaling dimension $\d[e]$. The blobs represent the rest of the diagram with arbitrary number of vertices, internal and external lines.\footnote{The blobs can have vertices in common. There can be also be other internal lines connecting $y_1$ and $y_2$, and in that case, we repeat the argument for each such internal line.} The figure on the right represents the same diagram but with the particular internal line cut at a point $z$, and with the scaling dimensions of the two new propagators so introduced given by $\Delta_c$. This introduces an extra vertex at $z$ that is integrated over all of spacetime. To see the equality of the diagrams depicted in Fig. \ref{cutting1}, we only have to perform the integral over $z$ on the right hand side, and show that it reproduces the propagator corresponding to the internal line of the diagram on the left hand side. This $z$ integral is given by
\begin{equation}\label{cuttingrule}
{1 \over \pi^{d/2}} \int d^dz {\Gamma(\d[c]) \over ((y_1-z)^2)^{\d[c]}} {\Gamma(\d[c]) \over ((y_2-z)^2)^{\d[c]}} = a_{\d[c]}{\Gamma(2\d[c]-d/2) \over ((y_1-y_2)^2)^{2\d[c]-d/2}},
\end{equation}
thus reproducing the propagator of the internal line of the diagram on the left hand side for the choice 
\begin{equation} \label{cutdelta}
	\d[c]={\d[e]+d/2 \over 2},
\end{equation}
up to the numerical constant $a_{\d[c]}=\frac{\Gamma^2(d/2-\d[c]) }{\Gamma(d-2\d[c])}$.
We now use this simple cutting rule and the proof of \eqref{flatfromsoft} for all contact contributions to provide a proof of the corresponding equality for all Feynman diagrams.

\subsection{Proof of equality of all Feynman diagrams} \label{allfeyneq}
Consider a general Feynman diagram with the number of vertices being $V$, and the number of internal lines being $I$. We use the identity in \eqref{cuttingrule} to cut every internal line. This replaces the propagator of the internal line with two propagators whose scaling dimensions are related to the scaling dimension of the original propagator by the relation \eqref{cutdelta}. Following this procedure reduces any general Feynman diagram to a product of $V$ contact diagrams with spacetime integrals over $I$ of the points.  Using our proof of the identity \eqref{flatfromsoft} for all contact diagrams, we can replace each such diagram by its conformal counterpart. Finally, we perform the spacetime integrals over all the intermediate points $z_i$ to find the correspondence \eqref{softarbitfeyn} relating Feynman diagrams with massless fields and their conformal counterparts.
 
We now formalize the argument of the previous paragraph. Any arbitrary Feynman diagram can be represented by the following expression
\begin{equation} \label{arbitfeyn}
	\begin{split}
		\mathcal{F}=\prod_{i=1}^{V} \left(\int d^dy_i \prod_{j \in E_i}\left(\frac{\Gamma(\Delta_j)}{((x_j-y_i)^2)^{\Delta_j}}\right) \prod_{k\in I'_i} \left(\frac{\Gamma(\Delta_k)}{((y_k-y_i)^2)^{\Delta_k}}\right)\right)
	\end{split}
\end{equation} 
where $i$ runs over all the vertices ranging from $1$ to $V$. For each vertex $i$, $j$ takes values in the set $E_i$ of indices of the external fields that connect to the vertex $i$. Let the set of indices of internal lines that connect to the vertex $i$ be $I_i$.  We assume an ordering of the vertices and let $k$ takes values in the set $I'_k$, the set of indices of internal lines at each vertex $k$ that haven't already been accounted for in the sets $I'_i$ with $i <k$. Now, we use the identity \eqref{cuttingrule} to replace each internal line in \eqref{arbitfeyn} by a spacetime integral over a product of propagators to find
 \begin{equation}
 	\begin{split}
 		\mathcal{F}
 		&= \prod_{i=1}^{V}\prod_{k\in I_i} {1 \over \pi^{d/2}} \int \frac{d^dz_{k}}{a_{\d[k]'}}     \int d^dy_i \prod_{j \in E_i}\left(\frac{\Gamma(\Delta_j)}{((x_j-y_i)^2)^{\Delta_j}}\right) {\Gamma(\Delta'_k) \over ((y_i-z_{k})^2)^{\Delta'_k}}
 	\end{split}
 \end{equation} 
where  $\Delta'_k=\half (\Delta_k+{d\over2})$, and we introduce the compact notation $\{\d[E_i]\}$ to mean the set of all scaling dimensions of external fields that connect to the vertex $i$, $\{\d[I_i]\}$ to mean the set of all scaling dimensions for internal fields introduced by the procedure of cutting, $\{x_{E_i}\}$ to mean the set of spacetime points of all external fields that connect to the vertex $i$, and finally $\{z_{I_i}\}$ for the set of spacetime points that are introduced by the cutting procedure for each of the internal lines that connect to the vertex $i$. Performing all the integrals over the $V$ vertices of the original diagram, we find that the expression for this diagram becomes a product of contact diagrams with spacetime integrals over $I$ points,
\begin{equation}
	\begin{split}
		\mathcal{F}
		&=  \prod_{k\in I_i}\int \frac{d^dz_{k}}{a_{\d[k]'}}  \prod_{i=1}^{V} \gflat_{\{\d[E_i]\},\{\d[I_i]\}}\left(\{x_{E_i}\},\{z_{I_i}\}\right)
	\end{split}
\end{equation}
  We now use our result for the equality of contact diagrams in \eqref{flatfromsoft} to convert each massless contact correlator to a conformal one with one extra insertion so that
  \begin{equation}
  	\begin{split}
  		\mathcal{F}=& \prod_{k\in I_i} \int \frac{d^dz_{k}}{a_{\d[k]'}}  \prod_{i=1}^{V} \lim_{x_{p_i}^2\rightarrow\infty} \frac{(x_{p_i}^2)^{\Dv[i]}}{\Gamma(\Dv[i])} \gcft_{\{\d[E_i]\},\Dv[i],\{\d[I_i]\}}\left(\{x_{E_i}\},x_{p_i},\{z_{I_i}\}\right)
  	\end{split}
  \end{equation} 
where we have defined $\Dv[i]=d-\sum_{i\in E_i} \d[i]-\sum_{i\in I_i} \d[i]$, so that the sum of scaling dimensions at each vertex now satisfies the conformal constraint \eqref{confconstraint}. We now use the cutting rule in \eqref{cuttingrule} again to perform all the $z$ integrals and find that 
\begin{equation} \label{arbitfeynconf}
	\begin{split}
		\mathcal{F}=\prod_{i=1}^{V}\lim_{x_{p_i}^2\rightarrow\infty} \frac{(x_{p_i}^2)^{\Dv[i]}}{\Gamma(\Dv[i])} \left(\int d^dy_i \prod_{j \in E_i}\left(\frac{\Gamma(\Delta_j)}{((x_j-y_i)^2)^{\Delta_j}}\right) \frac{\Gamma(\Dv[i])}{((x_{p_i}-y_i)^2)^{\Dv[i]}} \prod_{k\in I'_i} \left(\frac{\Gamma(\Delta_k)}{((y_k-y_i)^2)^{\Delta_j}}\right)\right)
	\end{split}
\end{equation} 
which is the conformal counterpart of \eqref{arbitfeyn} that we are after. Let us denote the contribution of \eqref{arbitfeyn} to the massless $n-$point correlator with $V$ vertices by $\gflat_{n,V}(x_1,\ldots,x_n)$. Let us also denote the conformal counterpart of this contribution with $n$ points and $n+V$ vertices by $\gcft_{n,n+V}(x_1,\ldots,x_{n+V})$. Then, we have shown the following result
\begin{equation}\label{softarbitfeyn}
	\begin{split}
		\gflat_{n,V}(x_1,\ldots,x_{n})= \prod_{i=1}^{V}\lim_{x_{p_i}^2\rightarrow\infty} \frac{(x_{p_i}^2)^{\Dv[i]}}{\Gamma(\Dv[i])} \gcft_{n,n+V}(x_1,\ldots,x_{n+V})
	\end{split}
\end{equation}
\section{Predictions for non-conformal correlators}\label{Exact}
In momentum space, Feynman diagrams get more complicated with the addition of a new momentum integral coming from each loop order. In position space, this complication comes from the addition of a spacetime integral for each new vertex. So position space correlators are naturally organized by the number of vertices in the diagrams. In this section, we use \eqref{softarbitfeyn} to find exact answers for some position space correlators.

We start with the simplest case of a single vertex. In \S \ref{sec:contact}, we saw that the answers for the 3-point and the 4-point non-conformal correlators can be computed and shown to be related to the 4-point and 5-point conformal correlators respectively. In \S \ref{5ptcont}, we write down the five-point contact correlator of massless fields. In \S \ref{2vert}, we will compute some contributions to 3-point and 4-point correlators coming from diagrams with two vertices, and in \S \ref{3vert}, we compute a contribution to the 3-point correlator from a diagram with three vertices. In \S \ref{nvert}, we compute some diagrams with an arbitrary number of vertices.
\subsection{5-point contact diagram} \label{5ptcont}

In this case, we will use our result to cast the five-point contact correlator as a conformal six-point correlator. Using \eqref{flatfromsoft}, we take the soft limit $x_6^2 \to \infty$ of the six-point conformal correlator in \eqref{gcft6} to write it as
\begin{equation} \label{gcft6}
	\gflata[1,5][1,5]={1\over \Gamma(d-\D[5])}{1 \over {  s_{12}^{\D[5]-d/2}}} \dflat_{\Delta_1,\ldots,\Delta_5}(r_1,\ldots,r_9)
\end{equation}
with 
\begin{equation} \label{dcft6}
	\begin{split}
	\dflat_{\Delta_1,\ldots,\Delta_5}(r_1,\ldots,r_9)
		= &\prod_{i=1}^{9}\int  {da_i \over {2\pi i}}   \,r_i^{a_i}  \,\Gamma(-a_i)  \Gamma(\Delta_5+a_{1234})\Gamma(\Delta_4+a_{4567})  \Gamma(\Delta_3+a_{3789}) \\
		&\times \Gamma\left({d \over2}-\Delta_{2345}-a_{234679}\right) \Gamma\left({d \over2}-\Delta_{1345}-a_{134578}\right) \Gamma\left(\D[5]-d/2-\A[9]\right),
	\end{split}	
\end{equation}
a function of the simple ratios 
\begin{equation} \label{cr6}
	\begin{split}
		&r_1=\frac{s_{15}}{s_{12}}, r_2=\frac{s_{25}}{s_{12}},r_3=\frac{s_{35}}{s_{12}},r_4=\frac{s_{45}}{s_{12}},\\
		&r_5=\frac{s_{14}}{s_{12}},	r_6=\frac{s_{24}}{s_{12}}, r_7=\frac{s_{34}}{s_{12}},\\
		&r_8=\frac{s_{13}}{s_{12}},r_9=\frac{s_{23}}{s_{12}}
	\end{split}
\end{equation}
obtained by taking the  $x_6^2 \to \infty$ of the conformal cross ratios for six points listed in \eqref{cr6}.

\subsection{Diagrams with two vertices} \label{2vert}
In \S \ref{1exch4p}, we provide the 1-exchange contribution to the 4-point correlator in Fig. \ref{4pt1exch} by casting it in terms of the conformal 6-point correlator in Fig. \ref{6pt1exch}. This easily generalizes to the 1-loop diagram for the 4-point correlator in Fig. \ref{4pt1l} as well as its $n-$loop versions with the same number of vertices. In \S \ref{1l3p}, we derive a 1-loop contribution to the 3-point correlator coming from Fig. \ref{3pt1l} by expressing it in terms of the 1-loop contribution to the conformal 6-point correlator given in Fig. \ref{5pt1l}. 

\subsubsection{1 -exchange contribution to the 4-point correlator} \label{1exch4p}
\begin{figure}
	\begin{center}
		\includegraphics[height=0.2\textheight]{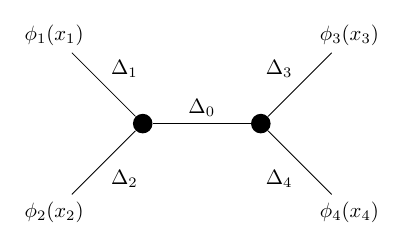}
		\caption{1-exchange contribution to the 4-point function}  \label{4pt1exch}
		
			\includegraphics[height=0.2\textheight]{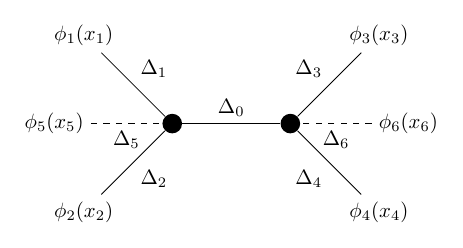}
		\caption{1-exchange contribution to the conformal 6-point function. The soft limit of this contribution, obtained by taking the dashed lines to infinity, gives the non-conformal 1-exchange contribution of Fig. \ref{4pt1exch}.}  \label{6pt1exch}
	\end{center}
\end{figure}

The simplest exchange diagram shown in Figure \ref{4pt1exch} appears at the second order in the perturbative expansion of the four-point correlator, given by 

\begin{equation} \label{gflate4o}
	\gflate4[\Delta_0]={1 \over \pi^d}\int d^d y_1 \left(\prod_{i=1}^{2}\frac{\Gamma(\d[i])}{((x_i-y_1)^2)^{\Delta_i}}\right) \int d^d y_2 \left(\prod_{i=3}^{4}\frac{\Gamma(\d[i])}{((x_i-y_2)^2)^{\Delta_i}} \right) \frac{\Gamma(\Delta_0)}{((y_1-y_2)^2)^{\Delta_0}}   
\end{equation}
Using the result \eqref{soft1exch}, we express this as the double soft limit of the conformal 6-point correlator of Fig. \ref{6pt1exch} whose value is given in \eqref{gcfte6} and find that
\begin{equation} \label{gflate4}
	\gflate4[\Delta_0] =\Gamma(d/2-\Delta_0)  \left(\frac{s_{14}s_{24}}{s_{12}}\right)^{d-\Delta_{04}}  \frac{1}{\prod_{i=1}^{3}s_{i4}^{\Delta_i}}   \dflate4
\end{equation}
\begin{equation} \label{dflate4}
	\begin{split}
		\dflate4
		= &\prod_{i=1}^{8}\int  {da_i \over {2\pi i}}   \ r_1^{a_1} r_2^{a_{23468}} r_3^{a_2} r_4^{a_{13457}} r_5^{-a_{1234}} \,\Gamma(-a_i)  \Gamma(\Delta_3+a_{1234})\Gamma(d-\Delta_{034}+a_{456}) \\ &\times\Gamma(d-\Delta_{012}+a_{378})
		\Gamma\left(d-\Delta_{04}+\A[8]\right)  \Gamma\left(\Delta_{14}-d-a_{23468}\right) \Gamma\left(\Delta_{24}-d-a_{13457}\right)\\
		& \times \frac{\Gamma(d/2-\Delta_{4}+a_{123456})}{\Gamma(d-\Delta_4+a_{123456})}
	\end{split}	
\end{equation}
a function of the simple ratios 
\begin{equation} \label{sr5}
	\begin{split}
		&r_1=\frac{s_{13}}{s_{12}}, r_2=\frac{s_{14}}{s_{12}},r_3=\frac{s_{23}}{s_{12}},r_4=\frac{s_{24}}{s_{12}},r_5=\frac{s_{34}}{s_{12}}
	\end{split}
\end{equation}
\paragraph{1-loop bubble diagram \\}

\begin{figure}
	\begin{center}
		\includegraphics[height=0.2\textheight]{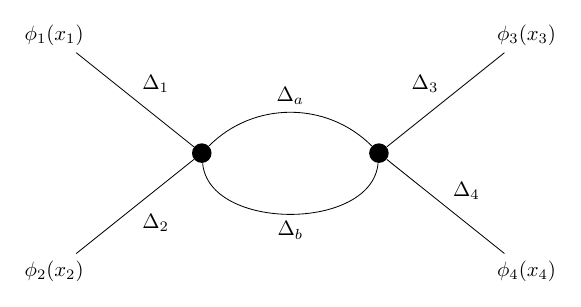}
		\caption{1-loop contribution to the 4-point function}  \label{4pt1l}
	\end{center}
\end{figure}

The 1-loop contribution to the 4-point function with two 4-point vertices given in Figure \ref{4pt1l} is

\begin{equation} \label{gflatl4}
	{\gflat}_{\ell,4} ={1 \over \pi^d}\int d^d y_1 \left(\prod_{i=1}^{2}\frac{\Gamma(\d[i])}{((x_i-y_1)^2)^{\Delta_i}}\right) \int d^d y_2 \left(\prod_{i=3}^{4}\frac{\Gamma(\d[i])}{((x_i-y_2)^2)^{\Delta_i}} \right) \frac{\Gamma(\Delta_{a})\Gamma(\Delta_{b})}{((y_1-y_2)^2)^{\Delta_{a}+\Delta_{b}}}   
\end{equation}
Apart from an overall numerical factor, the only difference from the 1-exchange contribution to the 4-point function with two 3-point vertices given in \eqref{gflate4o} is the change in power of the internal propagator. This readily gives us the answer as
\begin{equation} \label{gflatl4}
	{\gflat}_{\ell,4} =\frac{\Gamma(\d[a])\Gamma(\d[b])}{\Gamma(\d[ab])} \gflate4[\Delta_{ab}]
\end{equation}
in terms of the $\gflat$ function of the 1-exchange diagram given in \eqref{gflate4}

\paragraph{n-loop bubble diagram\\}
\begin{figure}
	\begin{center}
		\includegraphics[height=0.2\textheight]{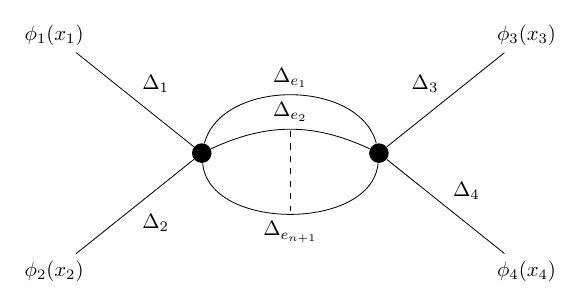}
		\caption{n-loop bubble contribution to the 4-point function}  \label{4ptnloop}
	\end{center}
\end{figure}

The $n-$loop bubble diagram in Fig. \ref{4ptnloop} is very simply obtained from the 1-exchange contribution in \eqref{gflate4o}. Denoting the scaling dimensions of the fields in the loop by $\d[e_1],\ldots,\d[e_{n+1}]$ and their sum by $\D[e_{n+1}]=\sum_{i=1}^{n+1} \d[e_i]$, we have that the $n-$loop bubble diagram is given by  
\begin{equation} \label{gflatnl4}
		{\gflat}_{n \ell,4} =\frac{\prod_{i=1}^{n+1} \Gamma(\d[e_i])}{\Gamma(\D[e_{n+1}])} \gflate4[\Delta^{(e_{n+1})}]
	\end{equation}

\subsubsection{1-loop contribution to the 3-point function} \label{1l3p}
\begin{figure}
	\begin{center}
		\includegraphics[height=0.2\textheight]{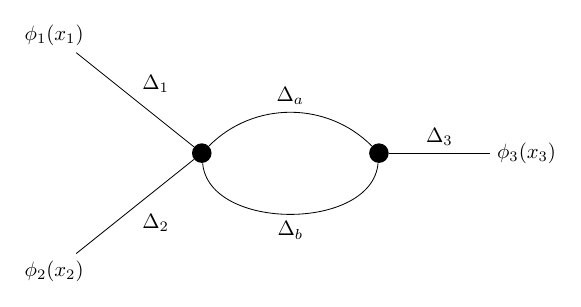}
		\caption{1-loop contribution to the 3-point function}  \label{3pt1l}
		\includegraphics[height=0.2\textheight]{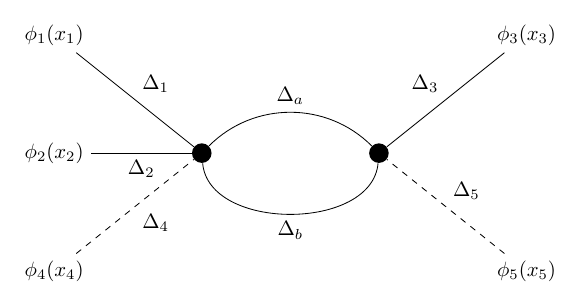}
		\caption{1-loop contribution to the conformal 5-point function. Taking the double soft limit corresponding to the fields with dashed lines reproduces the answer for the Feynman diagram in Fig. \ref{3pt1l}.}  \label{5pt1l}
	\end{center}
\end{figure}

The simplest 1-loop contribution to the 3-point function comes from the diagram with one cubic and one quartic vertex shown in Figure \ref{3pt1l}, and given by the expression

\begin{equation} \label{gflatl3}
	{\gflat}_{\ell,3} ={1 \over \pi^{d}}\int d^d y_1 \left(\prod_{i=1}^{2}\frac{\Gamma(\d[i])}{((x_i-y_1)^2)^{\Delta_i}}\right) \int d^d y_2 \frac{\Gamma(\d[3])}{((x_3-y_2)^2)^{\Delta_3}}  \frac{\Gamma(\Delta_{a})\Gamma(\Delta_{b})}{((y_1-y_2)^2)^{\Delta_{a}+\Delta_{b}}}   
\end{equation}

To convert this into a conformal correlator, we need to add an extra field at each vertex so that we get the diagram in Figure. \ref{5pt1l} given by
\begin{equation} 
	{\gcft}_{\ell,5} ={1 \over \pi^{d}}\int d^d y_1 \left(\prod_{i=1,2,4}\frac{\Gamma(\d[i])}{((x_i-y_1)^2)^{\Delta_i}}\right) \int d^d y_2 \left(\prod_{i=3,5}\frac{\Gamma(\d[i])}{((x_i-y_2)^2)^{\Delta_i}} \right) \frac{\Gamma(\Delta_{a})\Gamma(\Delta_{b})}{((y_1-y_2)^2)^{\Delta_{a}+\Delta_{b}}}   
\end{equation}
The scaling dimensions of the two extra operators $\Delta_4$ and $\Delta_5$ are chosen such that the conformal conditions at each vertex are satisfied
\begin{equation}
	\begin{split}
		\Delta_4=d-\Delta_{12}-\Delta_{ab} \\
		\Delta_5=d-\Delta_3-\Delta_{ab}
	\end{split}
\end{equation}
Now, we observe that, despite appearances, the $y_2$ integral is actually of the conformal 3-point contact type treated in Appendix \ref{cft3pt}. Using the result \eqref{cft3ptcont}, we find 
\begin{equation} 
	\begin{split}
		{\gcft}_{\ell,5} =&{1 \over \pi^{d/2}}\frac{\Gamma(\Delta_{a})\Gamma(\Delta_{b})}{\Gamma(\Delta_{ab})}\frac{\Gamma(\half \Delta_{35,ab})}{((x_3-x_5)^2)^{\half \Delta_{35,ab}}}    \int d^d y_1 \left(\prod_{i=1}^{5}\frac{\Gamma(\d[i'])}{((x_i-y_1)^2)^{\Delta_i'}}\right)  
	\end{split}
\end{equation}

where we have defined
\begin{equation}
	\Delta_i'=
	\begin{cases}
		\Delta_i   & i=1,2,4 \\
		\half \Delta_{3ab,5} & i=3 \\
		\half \Delta_{5ab,3} &  i=5 
	\end{cases}
\end{equation}
The integral over $y_1$ is now seen to be the conformal 5-point contact type. Using the results in Appendix \ref{cft5pt}, we find
\begin{equation} \label{gcftl5}
	\begin{split}
		{\gcft}_{\ell,5} =&\frac{\Gamma(\Delta_{a})\Gamma(\Delta_{b})}{\Gamma(\Delta_{ab})}\frac{\Gamma(\half \Delta_{35,ab})}{s_{35}^{\half \Delta_{35,ab}}}    \left(\frac{s_{12}}{s_{13}s_{23}}\right)^{\Delta_3'-d/2}  \frac{1}{\prod_{i=1,2,4,5}s_{i3}^{\Delta_i'}}\dcft_{\Delta'_1,\ldots,\Delta'_5}(s_1,\ldots,s_5)
	\end{split}
\end{equation}
where we have relabelled the field labels $3 \rightarrow 4, 4\rightarrow5,5\rightarrow3$ for convenience in taking the soft limits, so that
\begin{equation} 
	\begin{split}
		\dcft_{\Delta_1',\ldots,\Delta_5'}(s_1,\ldots,s_5)
		= &\prod_{i=1}^{5}\int  {da_i \over {2\pi i}}   \,s_i^{a_i}  \,\Gamma(-a_i)  \Gamma(\Delta'_4+a_{345})\Gamma(\Delta_5'+a_{123})   \\
		&\times \Gamma\left(\Delta'_{13}-a_{235}-{d \over2}\right)\Gamma\left(\Delta'_{23}-a_{134}-{d \over2}\right) \Gamma\left(\A[5]+{d \over2}-\Delta'_3\right)
	\end{split}	
\end{equation}
with the cross ratios given by
\begin{equation} \label{cr5p}
	s_1=\frac{s_{15}s_{23}}{s_{12}s_{53}}, s_2=\frac{s_{25}s_{13}}{s_{12}s_{53}},s_3=\frac{s_{13}s_{23}s_{45}}{s_{12}s_{43}s_{53}},s_4=\frac{s_{14}s_{23}}{s_{12}s_{43}},s_5=\frac{s_{24}s_{13}}{s_{43}s_{12}}
\end{equation}
Now, the 1-loop contribution to the 3-point function that we are after can be expressed in terms of the double soft limit of the 5-point conformal correlator above as follows
\begin{equation}
	\gflat_{\ell,3}=\lim\limits_{x_{4}^2,x_{5}^2\to \infty}  {(x_{4}^2)^{\Delta_4} (x_5^2)^{\Delta_5}\over \Gamma(\Delta_4) \Gamma(\Delta_5)}  \gcft_{\ell,5}. 
\end{equation}
Taking this double soft limit of \eqref{gcftl5}, we finally obtain
\begin{equation} \label{gflat3l}
	\begin{split}
		{\gflat}_{\ell,3} =&\frac{\Gamma(\Delta_{a})\Gamma(\Delta_{b})}{\Gamma(\Delta_{ab})} \frac{\Gamma(d/2-\Delta_{ab})}{\Gamma(d-\Delta_{12ab}) \Gamma(d-\Delta_{3ab})}    \left(\frac{s_{12}}{s_{13}s_{23}}\right)^{\Delta_{3ab}-d}  \frac{1}{\prod_{i=1,2}s_{i3}^{\Delta_i}}\dflat_{\Delta_1,\ldots,\Delta_3;\d[ab]}(r_1,r_2)
	\end{split}
\end{equation}
with 
\begin{equation} \label{dflat3l}
	\begin{split}
		\dflat_{\Delta_1,\ldots,\Delta_3;\d[ab]}(r_1,r_2)
		= &\prod_{i=1}^{4}\int  {da_i \over {2\pi i}}   \,r_1^{a_{13}} \, r_2^{a_{24}}  \,\Gamma(-a_i)  \Gamma(d-\Delta_{12ab}+a_{34})\Gamma\left({d \over 2}-\Delta_{3}+a_{12}\right)   \\
		&\times \Gamma\left(\Delta_{13ab}-{d \over 2}-a_{24} \right)\Gamma\left(\Delta_{23ab}-{d \over 2}-a_{13} \right) \Gamma\left(a_{1234}+d-\Delta_{3ab}\right)
	\end{split}	
\end{equation}
being a function of the two simple ratios
\begin{equation}
	r_1=\frac{s_{23}}{s_{12}}, r_2=\frac{s_{13}}{s_{12}}
\end{equation}
obtained by taking $x_4,x_5 \rightarrow \infty$ in the 5-point cross ratios given in \eqref{cr5p}.

We can easily generalize this answer to the 3-point correlator with two vertices and $n-$loops in place of the 1-loop in Fig. \ref{3pt1l}. Denoting the scaling dimensions of the fields in the loops by $\d[e_1],\ldots,\d[e_{n+1}]$, we have such an $n-$loop contribution (the analogue of Fig. \ref{4ptnloop}) to be given by
\begin{equation} 
	\begin{split}
		{\gflat}_{n\ell,3} =&\frac{\prod_{i=1}^{n+1}\Gamma(\Delta_{e_i})}{\Gamma(\D[e_{n+1}])} \frac{\Gamma(d/2-\D[e_{n+1}])}{\Gamma(d-\Delta_{12}-\D[e_{n+1}]) \Gamma(d-\Delta_{3}-\D[e_{n+1}])} \\
		&\times   \left(\frac{s_{12}}{s_{13}s_{23}}\right)^{\Delta_3+\D[e_{n+1}]-d}  \frac{1}{\prod_{i=1,2}s_{i3}^{\Delta_i}}\dflat_{\Delta_1,\ldots,\Delta_3;\D[e_{n+1}]}(r_1,r_2)
	\end{split}
\end{equation}
with $\dflat_{\Delta_1,\ldots,\Delta_3;\D[e_{n+1}]}(r_1,r_2)$ given by $\eqref{dflat3l}$ with $\d[ab]$ there replaced by $\D[e_{n+1}]$.
 
\subsection{Diagrams with three vertices} \label{3vert}
We move on to diagrams with three vertices which contribute to the 3-point massless correlator.

\subsubsection{2-loop contribution to the 3-point function}
\begin{figure}
	\begin{center}
		\includegraphics[height=0.2\textheight]{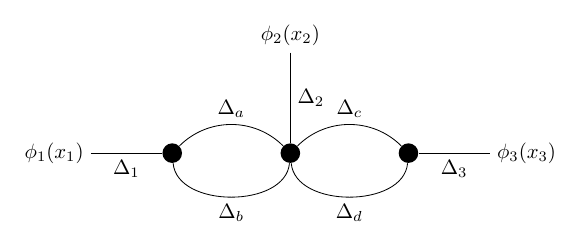}
		\caption{A 2-loop contribution to the 3-point function}  \label{3pt2l}
		\includegraphics[height=0.2\textheight]{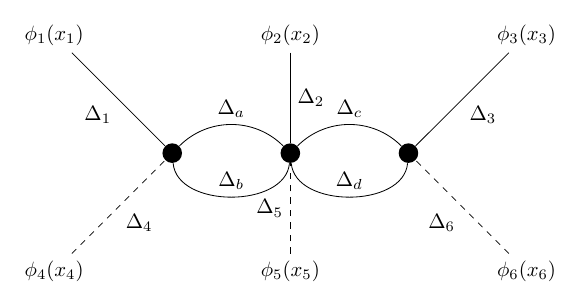}
		\caption{A 2-loop contribution to the conformal 6-point function. In this case, we take the triple soft limit by taking the dashed lines to infinity and find that the answer is the 2-loop contribution given in Fig. \ref{3pt2l}.}  \label{6pt2l1}
	\end{center}
\end{figure}

We compute the 2-loop Feynman diagram displayed in Figure \ref{3pt2l} which gives the expression
\begin{equation} \label{gflat2l31}
	{\gflat}_{2\ell,3} ={1 \over \pi^{3d/2}} \left(\prod_{i=1}^{3}\int d^d y_i\frac{\Gamma(\d[i])}{((x_i-y_i)^2)^{\Delta_i}}\right) \frac{\Gamma(\Delta_{a})\Gamma(\Delta_{b})}{((y_1-y_2)^2)^{\Delta_{a}+\Delta_{b}}} \frac{\Gamma(\Delta_{c})\Gamma(\Delta_{d})}{((y_2-y_3)^2)^{\Delta_{c}+\Delta_{d}}}   
\end{equation}
We will compute the answer to this integral by taking the triple soft limit of the conformal correlator contribution given by the Feynman diagram in Figure \ref{6pt2l1}.
The contribution of this Feynman diagram to the conformal 6-point function is
\begin{equation} 
	\begin{split}
		{\gcft}_{2\ell,6} =&{1 \over \pi^{3d/2}} \left(\prod_{i=1}^{3}\int d^d y_i\frac{\Gamma(\d[i])}{((x_i-y_i)^2)^{\Delta_i}}\right) \left(\prod_{i=1}^{3}\frac{\Gamma(\d[i+3])}{((x_{i+3}-y_i)^2)^{\Delta_{i+3}}}\right) \\ & \times \frac{\Gamma(\Delta_{a})\Gamma(\Delta_{b})}{((y_1-y_2)^2)^{\Delta_{a}+\Delta_{b}}} \frac{\Gamma(\Delta_{c})\Gamma(\Delta_{d})}{((y_2-y_3)^2)^{\Delta_{c}+\Delta_{d}}}  
	\end{split}
\end{equation}
This time, both the $y_1$ and $y_3$ integrals are of the 3-point conformal contact type, and can be done using the result of Appendix \ref{cft3pt} to find
\begin{equation} 
	\begin{split}
		{\gcft}_{2\ell,6} =&{1 \over \pi^{d/2}}\frac{\Gamma(\Delta_{a})\Gamma(\Delta_{b})}{\Gamma(\Delta_{ab})}\frac{\Gamma(\Delta_{c})\Gamma(\Delta_{d})}{\Gamma(\Delta_{cd})} \frac{\Gamma(\half \Delta_{14,ab})}{((x_1-x_4)^2)^{\half \Delta_{14,ab}}} \frac{\Gamma(\half \Delta_{36,cd})}{((x_3-x_6)^2)^{\half \Delta_{36,cd}}}    \\
		& \times \int d^d y_2 \left(\prod_{i=1}^{6}\frac{\Gamma(\d[i'])}{((x_i-y_2)^2)^{\Delta_i'}}\right)  
	\end{split}
\end{equation}

with the definition
\begin{equation}
	\Delta_i'=
	\begin{cases}
		\Delta_i   & i=2,5 \\
		\half \Delta_{1ab,4} & i=1 \\
		\half \Delta_{4ab,1} &  i=4 \\ 
		\half \Delta_{3cd,6} &  i=3 \\ 
		\half \Delta_{6cd,3} &  i=6 \\ 
	\end{cases}
\end{equation}

The $y_2$ integral is of the conformal 6-point contact type treated in Appendix \ref{cft6pt}. Using the result \eqref{gcft6} given there (and after relabelling 6 $\leftrightarrow$ 3), we find
\begin{equation} \label{gcft2l6}
	\begin{split}
		{\gcft}_{2\ell,6} =&\frac{\Gamma(\Delta_{a})\Gamma(\Delta_{b})}{\Gamma(\Delta_{ab})}\frac{\Gamma(\Delta_{c})\Gamma(\Delta_{d})}{\Gamma(\Delta_{cd})} \frac{\Gamma(\half \Delta_{14,ab})}{s_{14}^{\half \Delta_{14,ab}}} \frac{\Gamma(\half \Delta_{36,cd})}{s_{36}^{\half \Delta_{36,cd}}} \\
		& \times \left(\frac{s_{12}}{s_{13}s_{23}}\right)^{\Delta_3'-d/2}  \frac{1}{\prod_{i=1,2,4,5,6}s_{i3}^{\Delta_i'}}\dcft_{\Delta'_1,\ldots,\Delta'_6}(s_1,\ldots,s_9)
	\end{split}
\end{equation}
where
\begin{equation}
	\begin{split}
		\dcft_{\Delta'_1,\ldots,\Delta'_6}(s_1,\ldots,s_9)
		= &\prod_{i=1}^{9}\int  {da_i \over {2\pi i}}   \,s_i^{a_i}  \,\Gamma(-a_i)  \Gamma(\Delta_5+a_{1234})\Gamma(\Delta'_4+a_{4567})  \Gamma(\Delta'_6+a_{3789}) \\
		&\times \Gamma\left(\Delta'_{13}-{d \over2}-a_{234679}\right) \Gamma\left(\Delta'_{23}-{d \over2}-a_{134578}\right) \Gamma\left({d \over2}-\Delta'_3-\A[9]\right)
	\end{split}	
\end{equation}
is a function of the cross ratios 
\begin{equation} \label{cr6p}
	\begin{split}
		&s_1=\frac{s_{15}s_{23}}{s_{12}s_{35}}, s_2=\frac{s_{25}s_{13}}{s_{12}s_{35}},s_3=\frac{s_{56}s_{13}s_{23}}{s_{12}s_{36}s_{35}},s_4=\frac{s_{45}s_{13}s_{23}}{s_{12}s_{34}s_{35}},\\
		&s_5=\frac{s_{14}s_{23}}{s_{12}s_{34}},	s_6=\frac{s_{24}s_{13}}{s_{12}s_{34}}, s_7=\frac{s_{46}s_{13}s_{23}}{s_{12}s_{36}s_{34}},\\
		&s_8=\frac{s_{16}s_{23}}{s_{12}s_{36}},s_9=\frac{s_{26}s_{13}}{s_{36}s_{12}}
	\end{split}
\end{equation}

The 2-loop contribution to the 3-point function \eqref{gflat2l31} can be found by taking the triple soft limit of the 6-point conformal correlator above as
\begin{equation}
	\gflat_{2\ell,3}=\lim\limits_{x_{4}^2,x_{5}^2,x_{6}^2\to \infty}  {(x_{4}^2)^{\Delta_4} (x_5^2)^{\Delta_5} (x_6^2)^{\Delta_6}\over \Gamma(\Delta_4) \Gamma(\Delta_5) \Gamma(\Delta_6)}  \gcft_{2\ell,6}. 
\end{equation}
Taking this triple soft limit of \eqref{gcft2l6}, we finally obtain
\begin{equation} 
	\begin{split}
		{\gflat}_{2\ell,3} =& \frac{\Gamma(\Delta_{a})\Gamma(\Delta_{b})}{\Gamma(\Delta_{ab})}\frac{\Gamma(\Delta_{c})\Gamma(\Delta_{d})}{\Gamma(\Delta_{cd})} \frac{\Gamma(d/2- \Delta_{ab}))\Gamma(d/2- \Delta_{cd})}{\Gamma(d-\Delta_{1ab}) \Gamma(d-\Delta_{3cd}) \Gamma(d-\Delta_{2abcd})} \\
		& \left(\frac{s_{12}}{s_{13}s_{23}}\right)^{\Delta_{3cd}-d/2}  \frac{1}{\prod_{i=1,2}s_{i3}^{\Delta_i'}}\dflat_{\Delta_1,\ldots,\Delta_3;\d[ab],\d[cd]}(r_1,r_2)
	\end{split}
\end{equation}
with 
\begin{equation} \label{dflat32l1}
	\begin{split}
		\dflat_{\Delta_1,\ldots,\Delta_3;\d[ab],\d[cd]}(r_1,r_2)
		= &\prod_{i=1}^6\int  {da_i \over {2\pi i}}   \,r_1^{a_{135}} \,r_2^{a_{246}}\,\Gamma(-a_i)  \Gamma(d-\Delta_{2abcd}+a_{12})\Gamma({d \over2}-\Delta_{1}+a_{34}) \\
		&\times \Gamma({d \over2}-\Delta_{3}+a_{56})  \Gamma\left(\Delta_{1ab,3}-a_{246}\right) \Gamma\left({d \over2}+\Delta_{2,3}-a_{135}\right) \Gamma\left(d -\Delta_{3cd}-\A[6]\right)
	\end{split}	
\end{equation}
being a function of the two simple ratios
\begin{equation}
	r_1=\frac{s_{23}}{s_{12}}, r_2=\frac{s_{13}}{s_{12}}
\end{equation}
obtained by taking $x_4,x_5,x_6 \rightarrow \infty$ in the 6-point cross ratios given in \eqref{cr6p}.
\begin{figure}
	\begin{center}
		\includegraphics[height=0.2\textheight]{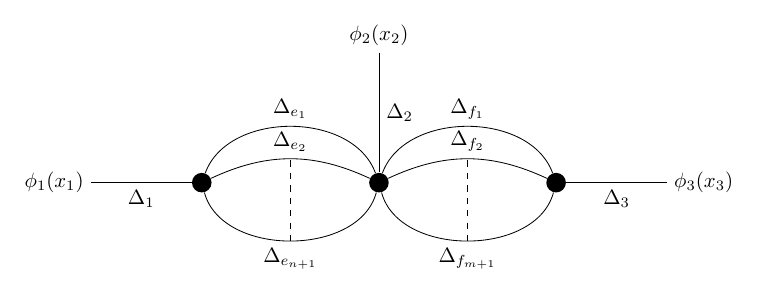}
		\caption{A $(n+m)$-loop contribution to the 3-point function}  \label{3ptnl1}
	\end{center}
\end{figure}
Just as before, we can easily generalize this answer to find the $(n+m)$ loop contribution to the 3-point correlator corresponding to the diagram Fig. \ref{3ptnl1}. With the assignment of the scaling dimensions as depicted in Fig. \ref{3ptnl1}, and using the notations $\D[f_{m+1}]=\sum_{i=1}^{m+1}\d[f_i]$ and  $\D[e_{n+1}+f_{m+1}]=\D[e_{n+1}]+\D[f_{m+1}]$, we find 
\begin{equation} 
	\begin{split}
		{\gflat}_{(n+m)\ell,3} =& \frac{\prod_{i=1}^{n+1}\Gamma(\Delta_{e_i})\prod_{i=1}^{m+1}\Gamma(\Delta_{f_i})}{\Gamma(\D[e_{n+1}])\Gamma(\D[f_{m+1}])}\frac{\Gamma(d/2- \D[e_{n+1}]))\Gamma(d/2- \D[f_{m+1}])}{\Gamma(d-\Delta_{1}-\D[e_{n+1}]) \Gamma(d-\Delta_{3}-\D[f_{m+1}]) \Gamma(d-\Delta_{2}-\D[e_{n+1}+f_{m+1}]} \\
		& \times \left(\frac{s_{12}}{s_{13}s_{23}}\right)^{\Delta_3-d/2}  \frac{1}{\prod_{i=1,2}s_{i3}^{\Delta_i'}}\dflat_{\Delta_1,\ldots,\Delta_3;\D[e_{n+1}],\D[f_{m+1}]}(r_1,r_2)
	\end{split}
\end{equation}
where $\dflat_{\Delta_1,\ldots,\Delta_3;\D[e_{n+1}],\D[f_{m+1}]}(r_1,r_2)$ is given by \eqref{dflat32l1} with $\d[ab]$ and $\d[cd]$ there replaced by $\D[e_{n+1}]$ and $\D[f_{m+1}]$ respectively.

\subsection{Diagrams with arbitrary number of vertices}\label{nvert}
\begin{figure}
	\begin{center}
		\includegraphics[height=0.2\textheight]{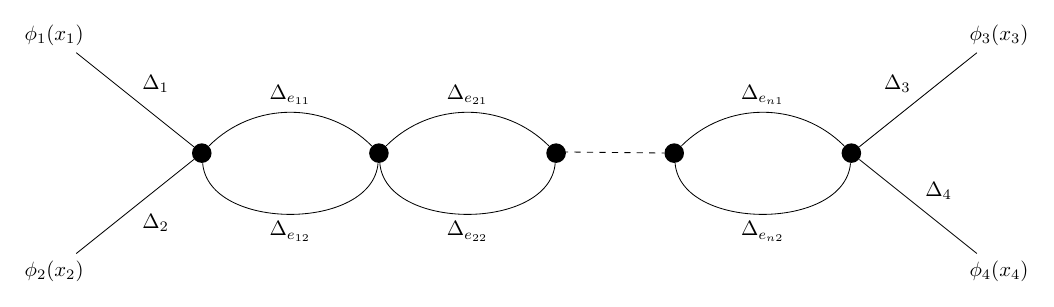}
		\caption{Contribution to the 4-point function from a bubble diagram with n-loops.}  \label{4ptnb}
	
	\end{center}
\end{figure}
The classes of diagrams studied in the last two subsections can be generalized to $n$ vertices. First, we consider contributions to the 4-point massless correlator given in Fig. \ref{4ptnb}. All the integrals over \emph{internal vertices} i.e. vertices with only internal lines joining them, can be easily done via use of the identity \eqref{cuttingrule} to sew these vertices together, two at a time. There are $(n-1)$ such internal vertices. We use the notation $\Delta_{e_k}=\Delta_{e_{k1}}+\Delta_{e_{k2}}$.  It is easy to see that, after applying \eqref{cuttingrule} $(n-1)$ times, this diagram reduces to the 1-exchange diagram of Fig. \ref{4pt1exch} with the scaling dimension of the exchanged field being given by $$\Delta_e\equiv\sum_{i=1}^{n} \Delta_{e_{i}}- (n-1)d/2=\D[e_n]-(n-1)d/2.$$
 In this way, we get for the diagram in Fig. \ref{4ptnb},
\begin{equation}
	\gflat_{nb,4}=\prod_{i=1}^n \prod_{j=1}^2 \frac{\Gamma(\d[e_{ij}]) \Gamma(d/2-\d[e_i])}{\Gamma(\d[e_i])} \gflate4[\Delta_e]
\end{equation}
in terms of the correlator for the 1-exchange diagram given in \eqref{gflate4}.

\begin{figure}
	\begin{center}
		\includegraphics[height=.18\textheight]{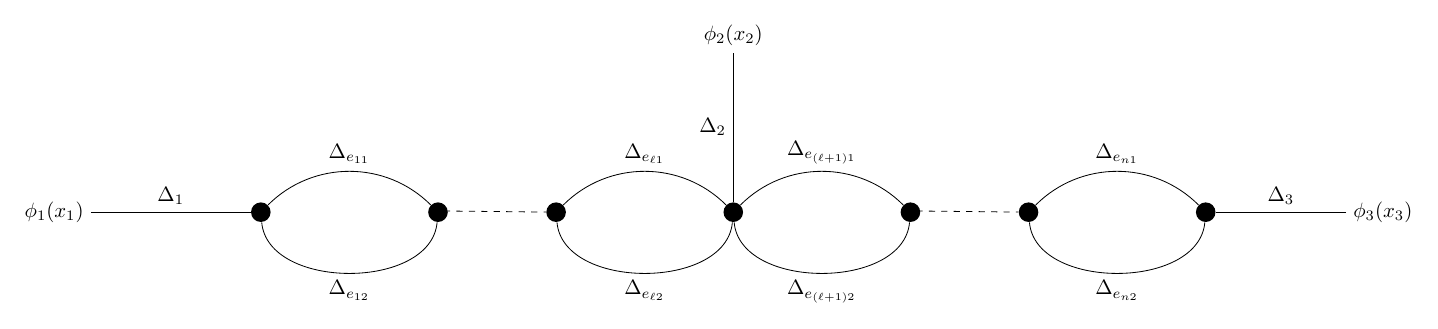}
		\caption{Contribution to the 3-point function from a bubble diagram with $n=(\ell+r)$-loops}  \label{3ptnb}		
	\end{center}
\end{figure}
Similarly, we can compute contributions to the $3-$point correlator coming from Fig. \ref{3ptnb} which has $l$ bubbles on the left and $r$ on the right, for a total of $n=\ell+r$ bubbles.
Using the cutting rule identity \eqref{cuttingrule} to sew together all the internal vertices, we end up with the 3-point contact diagram. For convenience, we define
\begin{equation}
	\d[\ell]=\sum_{i=1}^{n} \Delta_{e_{i}}- (\ell-1)d/2 , \quad \d[r]=\sum_{i=\ell+1}^{n} \Delta_{e_{i}}- (r-1)d/2	
\end{equation}
The contribution from Fig. \ref{3ptnb} is given by
\begin{equation} \label{gflat3nb}
			\gflat_{nb,3}=\prod_{i=1}^n \prod_{j=1}^2 \frac{\Gamma(\d[e_{ij}]) \Gamma(d/2-\d[e_i])}{\Gamma(\d[e_i])} a_{\d[1],\d[\ell]} a_{\d[3],\d[r]} \ \gflat_{\d[1\ell]-d/2,\d[2],\d[3 r]-d/2}(x_1,x_2,x_3)
\end{equation}
in terms of the 3-point contact $\gflat$ function given in \eqref{gflat3}\footnote{where we have defined $a_{\d[1],\d[2]}=\frac{\Gamma(d/2-\d[1]) \Gamma(d/2-\d[2])}{\Gamma(d-\d[12])}$, the generalization of the numerical constant in \eqref{cuttingrule}.}.

The 1-loop contribution to the 3-point function in Fig. \ref{3pt1l} is a special case of the diagram in Fig. \ref{3ptnb} i.e. with $\ell=0$ and $r=1$. This means that this 1-loop answer that we computed in \eqref{gflat3l} actually reduces to the simpler 3-point contact answer in \eqref{gflat3nb}. Similarly, the 2-loop contribution to the 3-point function in Fig. \ref{3pt2l} is a special case of Fig. \ref{3ptnb} with $\ell=1$ and $r=1$, thus reducing \eqref{gflat2l31} to the 3-point contact answer in \eqref{gflat3nb}. It would be interesting to explore these relations that reduce more complicated $\dflat$ functions to simpler ones.

\section{Correlators of massive fields}\label{massive}
\paragraph{Contact contributions\\}
The contact contribution to the $n$-point massive correlator coming from a $\lambda \phi^n$ type of interaction was expressed in terms of the $\dflat$ functions in Appendix H of \cite{Pibv}. We first review this construction and then use the results derived in this article to express the contact massive correlators as infinite sums of the kind of conformally invariant correlators treated here.

Just as in \S \ref{sec:contact}, we can express the $n-$point massive contact diagram as 

\begin{equation}
	\iflata[1,n][1,n]=\mathcal{A} \int d^d y \prod_{i=1}^{n} G(x_i-y)
\end{equation}
whereas the propagator is given by \footnote{Normalizing the propagator so that	$(\nabla^2-m^2) G(x_1- x_2)=\delta^{(d)}(x_1-x_2)$, introduces an overall constant factor given by  $ 2^{2 \Delta_i-d/2} m_i^{d}$. Instead, our normalization factor is chosen so that in the limit of vanishing masses of a massive correlator, we recover the corresponding massless correlator.}

\begin{equation}
	G(x_i-y)={1 \over \pi^{d/2}}\left({m_i^2 \over 4}\right)^{\Delta_i} \frac{K_{\Delta_i}(\sqrt{s_i})}{s_i^{\Delta_i/2}} 
\end{equation}
 with the definitions  $\Delta_i=\dm2$, $s_i=m_i^2 (x_i-y)^2$. Just as in the massless case, we drop the normalization constant $\mathcal{A}$ in what follows.

We now use the following integral representation for the modified Bessel function of the second kind that appears in all the propagators
\begin{equation}\label{ksch}
	K_{\nu}(z)=\half \left({z \over 2}\right)^{\nu} \int_0^{\infty} \frac{e^{-t-{z^2 \over 4 t}}}{t^{\nu+1}} dt
\end{equation}
 This converts the $y$ integral to a Gaussian, much like the Schwinger parametrization we used in the massless case. Performing the bulk integral and using the variable redefinition ${m_i^2 \over 4 t_i} \rightarrow t_i$, we find that
\begin{equation} 
	\begin{split}
		\iflata[1,n][1,n]&= \prod_i^n \int_{0}^{\infty} dt_i t_i^{\Delta_i-1} e^{- {m_i^2 \over 4 t_i} }{1 \over T^{d/2}} \exp(-{1\over T} \sum\limits_{\substack{i,j=1 \\ i \le j}}^{n} t_i t_j s_{ij}),
	\end{split}
\end{equation}
the analogue of \eqref{eq:schpara} in the massless case. In fact, the only difference from the massless case is the exponential term involving the masses. This means that the massive correlator, as represented above, smoothly approaches the corresponding massless correlator as the masses vanish. This fact is particularly convenient in the computation of correlators involving both massless and massive fields.

We can obtain an expansion of the massive correlator perturbatively in the masses of the particles by Taylor expanding the exponentials to find 
\begin{equation} 
	\begin{split}
			\iflata[1,n][1,n]&=\prod_{i=1}^n \sum_{p_i=0}^{\infty} {1 \over p_i!}\left(-{{m_i^2} \over {4 }}\right)^{p_i} \ \int_{0}^{\infty} dt_i t_i^{\Delta_i-1-p_i}{1 \over T^{d/2}} \exp(-{1\over T} \sum\limits_{\substack{i,j=1 \\ i \le j}}^{n} t_i t_j s_{ij}),
	\end{split}
\end{equation}
and identifying the coefficient at every perturbative order to be the massless correlator $\gflat$ of $n$ fields given in \eqref{eq:schpara} with scaling dimensions shifted by the order of perturbation,
\begin{equation} 
	\begin{split}
			\iflata[1,n][1,n]&=\prod_{i=1}^n\sum_{p_i=0}^{\infty} {1 \over p_i!} \left(-{{m_i^2} \over {4 }}\right)^{p_i} \ \gflat_{\d[1]-p_1,\ldots\d[n]-p_n}(x_1,\ldots,x_n)
	\end{split}
\end{equation}
 We now use \eqref{flatfromsoft} to express the massive contact correlator in terms of infinite sums of conformally invariant correlators as
 \begin{equation} 
 	\begin{split}
 		\iflata[1,n][1,n]&=\prod_{i=1}^n\sum_{p_i=0}^{\infty} \lim_{x_{n+1}^2\rightarrow\infty} {1 \over p_i!} \left(-{{m_i^2} \over {4 }}\right)^{p_i} \frac{(x_{n+1}^2)^{\d[n+1]}}{\Gamma(\d[n+1])} \  \gcft_{\d[1]-p_1,\ldots\d[n]-p_n,\d[n+1]}(x_1,\ldots,x_{n+1})
 	\end{split}
 \end{equation}
where $\d[n+1]=d+\sum_{i=1}^{n}(p_i-\d[i])$.

\paragraph{General Feynman diagrams\\}
We now show that the contribution to massive correlators coming from any Feynman diagram can be expressed as a infinite sum of conformally invariant correlators. Let us begin with the 1-exchange diagrams i.e the massive counterpart of \eqref{gflaten},
\begin{equation} \label{iflaten}
\begin{split}
		{\iflat}_{e,n} =&{1 \over \pi^d}\int d^d y_1 \prod_{i=1}^{\ell}\left({m_i \over 4}\right)^{\Delta_i} \frac{K_{\Delta_i}\left(m_i\sqrt{(x_i-y_1)^2}\right)}{\left({(x_i-y_1)^2}\right)^{\Delta_i/2}}  \int d^d y_2 \prod_{j=\ell+1}^{\ell+r}\left({m_i^2 \over 4}\right)^{\Delta_i} \frac{K_{\Delta_j}\left(m_j\sqrt{(x_j-y_2)^2}\right)}{\left({(x_j-y_2)^2}\right)^{\Delta_j/2}}   \\
		& \times \left({m_0^2 \over 4}\right)^{\Delta_0} \frac{K_{\Delta_0}\left(m_0\sqrt{(y_1-y_2)^2}\right)}{\left({(y_1-y_2)^2}\right)^{\Delta_0/2}} ,
\end{split}   
\end{equation}
Using the integral representation in \eqref{ksch} for all the fields connected to the vertex at $y_1$, and doing the bulk integral over $y_1$ we arrive at the massive analogue of \eqref{lint},
\begin{equation} \label{mlint}
	\begin{split}
		{\iflat}_{e,n} =&{1 \over \pi^{d/2}} \int d^d y_2 \left( \prod_{j=\ell+1}^{\ell+r}\left({m_i^2 \over 4}\right)^{\Delta_i} \frac{K_{\Delta_j}\left(m_j\sqrt{(x_j-y_2)^2}\right)}{\left({(x_j-y_2)^2}\right)^{\Delta_j/2}}  \right)   \\
		&\times \prod_{i=0}^{\ell} \int_{0}^{\infty} dt_i t_i^{\Delta_i-1}e^{- {m_i^2 \over 4 t_i} }{1 \over T_{L}^{d/2}} \exp(-{1\over T_{L}} \sum\limits_{\substack{i,j=0 \\ i \le j}}^{\ell} t_i t_j s_{ij}),
	\end{split}
\end{equation}
the only difference from the massless case being the exponential terms in the masses. Again series expanding these terms, we find
\begin{equation} \label{mlint2}
	\begin{split}
		{\iflat}_{e,n} =&{1 \over \pi^{d/2}} \int d^d y_2 \left( \prod_{j=\ell+1}^{\ell+r}\left({m_i^2 \over 4}\right)^{\Delta_i} \frac{K_{\Delta_j}\left(m_j\sqrt{(x_j-y_2)^2}\right)}{\left({(x_j-y_2)^2}\right)^{\Delta_j/2}}  \right)   \\
	& \times \prod_{i=0}^{\ell} \sum_{p_i=0}^{\infty} {1 \over p_i!}\left(-{{m_i^2} \over {4 }}\right)^{p_i}\int_{0}^{\infty} dt_i t_i^{\Delta_i-p_i-1}{1 \over T_{L}^{d/2}} \exp(-{1\over T_{L}} \sum\limits_{\substack{i,j=0 \\ i \le j}}^{\ell} t_i t_j s_{ij}),
	\end{split}
\end{equation}
We make the variable substitution of \eqref{schsub} and use the integral representation \eqref{ksch} for all the Bessel functions in \eqref{mlint2}. Performing the bulk integral over $y_1$, and series expanding the resultant exponential terms in masses, we obtain the analogue of \eqref{lrint} to be
\begin{equation} \label{mlrint}
	{\iflat}_{e,n} =   \prod_{i=0}^{n} \sum_{p_i=0}^{\infty} {1 \over p_i!}\left(-{{m_i^2} \over {4 }}\right)^{p_i} \int_{0}^{\infty} dt_i t_i^{\Delta_i-p_i-1}{1 \over T_{L}^{d-\Delta_L}}e^{- \sum\limits_{ L } t_i t_j s_{ij}}  {1 \over W^{d/2}}e^{- {1 \over W}\sum\limits_{L+R } w_i w_j s_{ij}}
\end{equation}
The rest of the treatment in \S \ref{exch} carries through and we finally find the analogue of \eqref{flat1e}
\begin{equation} \label{m1e}
	{\iflat}_{e,n} =   \prod_{i=0}^{n}  \sum_{p_i=0}^{\infty} {1 \over p_i!}\left(-{{m_i^2} \over {4 }}\right)^{p_i}\int_{0}^{\infty} da_i a_i^{\Delta'_i-1}{1 \over (a_0 A+A_{L'})^{d-\Delta'_L}}e^{- \sum\limits_{ L' } a_i a_j s_{ij}}  {1 \over A^{d-\Delta'_R}}e^{- \sum\limits_{L'+R'} a'_i a'_j s_{ij}}
\end{equation}
with all the scaling dimensions $\d[i]$ in \eqref{flat1e} replaced by $\d[i]'=\d[i]-p_i$.
This means that, after introducing all the labels to the 1-exchange contribution $\gflat_{e,n}$ by $\gflat_{\d[1],\ldots\d[n];\d[0]}(x_1,\ldots,x_n)$, we can write the 1-exchange contribution as
\begin{equation} 
	\begin{split}
		{\iflat}_{e,n}&=\prod_{i=1}^n\sum_{p_i=0}^{\infty} {1 \over p_i!} \left(-{{m_i^2} \over {4 }}\right)^{p_i} \ \gflat_{\d[1]-p_1,\ldots\d[n]-p_n;\d[0]-p_0}(x_1,\ldots,x_n)
	\end{split}
\end{equation}
Now using the result in \eqref{soft1exch} to write the massless 1-exchange $n-$point answer as the conformal $(n+2)-$ point diagram $\gcft_{e,n+2}$ which we denote here by $\gcft_{\d[1]-p_1,\ldots\d[n+2]-p_{n+2};\d[0]-p_0}(x_1,\ldots,x_{n+2})$ with scaling dimensions of the two additional fields given by $\d[\ell+1]=d-\Delta'_L=d-\Delta_L-\sum_{i=1}^{\ell} p_i$ and $\d[r+1]=d-\Delta'_R=d-\Delta_R-\sum_{i=\ell+1}^{n} p_i$, we find
\begin{equation} 
	\begin{split}
		{\iflat}_{e,n}
		&=\prod_{i=1}^n\sum_{p_i=0}^{\infty}\lim_{\substack{x_{\ell+1}^2\rightarrow\infty\\x_{r+1}^2\rightarrow\infty}}   {1 \over p_i!} \left(-{{m_i^2} \over {4 }}\right)^{p_i} \ \frac{(x_{\ell+1}^2)^{\d[\ell+1]}}{\Gamma(\d[\ell+1])} \frac{(x_{r+1}^2)^{\d[r+1]}}{\Gamma(\d[r+1])}  \gcft_{\d[1]-p_1,\ldots\d[n+2]-p_{n+2};\d[0]-p_0}(x_1,\ldots,x_{n+2})
	\end{split}
\end{equation}
In this way, we arrive at the result that all 1-exchange massive correlators can be expressed as infinite sums of the corresponding conformal correlators. It is not hard to see that the proofs of this section can be generalized to find such a representation for arbitrary Feynman diagrams. The proof for a general massless correlator can be uplifted to a proof for the corresponding massive correlator by replacing the introduction of every Schwinger parameter with the use of the integral representation in \eqref{ksch}. The only difference from the case of the massless correlator are exponential terms in masses, which we series expand, and then use the same manipulations as in the massless case for each term in the sum but with shifted scaling dimensions.

\section{Discussion}\label{disc}

Quantum field theory correlators in flat spacetimes are of course simpler in momentum space where they are usually studied. On the other hand, when studying conformal field theories, the additional symmetries allow for analysis in position space. In this article, we have expressed the Feynman diagram contributions to quantum field correlators as functions of only the conformal invariants, as if they were correlators coming from a conformal field theory. In other words, even while the theory is not conformal, we see that the correlators can be built out of conformally invariant $\dcft$ functions which are well studied in the  context of the AdS/CFT correspondence. Conformal symmetry ensures the existence of various relations between the resulting $\dcft$ functions, such as the recurrence relations of \cite{Dolan:2000ut}, using which it would be useful to understand the simplest representation for the $\dflat$ functions.

We note that the method of Mellin-Barnes representation used to evaluate the conformal correlators, as first done in \cite{Symanzik:1972wj} has been successfully and widely employed in the treatment of flat space correlators in momentum space starting with the work of \cite{Usyukina:1975yg,Bergere:1973fq} as reviewed, for example, in the books \cite{Smirnov:2006ry,Weinzierl:2022eaz}. In this approach, the appearance of conformally invariant functions is observed in some special cases. Recently, systematic algorithms for the evaluation of multiple Mellin-Barnes integrals have been proposed, and implemented in packages in \cite{Ananthanarayan:2020fhl,Banik:2022bmk,Banik:2023rrz}. These could be used to actually compute the different Mellin-Barnes integrals we encounter here. As we noted in \S\ref{sec:contact}, some of the conformally invariant functions we encounter also appear in the computation of conformal blocks \cite{Ferrara:1972kab,Ferrara:1974nf,Ferrara:1974ny,Dolan:2000ut,Dolan:2001,Dolan:2003hv,Dolan:2011dv}. It would be great to understand this connection better.

Studying correlators directly in position space is especially useful when one is interested in their behaviour  as the insertion points are taken to the boundary of the spacetime. In \cite{Laddha:2020kvp}, it was argued that in the case of massless fields, such boundary correlators defined on the past of null infinity contain all bulk information about massless excitations. It would be interesting to apply the methods presented here to study such correlators. Relatedly, in \cite{Pibv}, it was shown that the singularities of these correlators give bulk S-matrices. There, we start with the Euclidean correlators computed here, and perform an analytic continuation to the Lorentzian configuration. We see that the correlators don't have any singularities while on the Euclidean sheet, and develop singularities when continued to Lorentzian configuration precisely when the external insertions are such that they can scatter in the bulk in a momentum conserving manner. The residues at these singularities encode the bulk S-matrices in a manner reminiscent of what happens in CFTs (see \cite{Gary:2009ae,Chandorkar:2021viw}). For this reason, characterizing the singularity structure of the correlators is an important problem that we hope to address. Our results effectively reduce this problem to that of studying the singularities of CFT correlators. After performing all the Mellin integrals, these correlators can be written as generalized hypergeometric functions of several variables, and we would need to understand the nature of their singularities. 

We found that massive correlators can be written as an infinite sum over massless correlators, and hence also as an infinite sum over conformal correlators. It would be interesting to use this representation to find the extrapolate limit of these correlators as the insertion points are taken to spatial infinity, as was done in \cite{Laddha:2022nmj}. This could shed light on understanding how holography works for massive excitations in flat spacetimes, in particular the holography of information \cite{Laddha:2020kvp,Raju:2020smc,Chowdhury:2021nxw,Raju:2021lwh,Chowdhury:2022wcv,deMelloKoch:2022sul,Chakravarty:2023cll,Chakraborty:2023los}.

As we have mentioned, in the computation of the conformal correlators, we find the same conformally invariant integrals that appear when calculating correlators in AdS spacetimes \cite{DHoker:1999kzh,Dolan:2000ut}. This naturally leads to the question of what the AdS interpretation of our results is. In particular, this suggests a new way of computing the flat space correlators and amplitudes from AdS correlators. Studying the implications of such a limit is a useful question for future study. We have only considered scalar fields here, and a natural extension of our results is the generalization to include gauge and gravity fields. We would expect some of the $\dcft$ functions of \cite{DHoker:1999kzh,DHoker:1999mic,DHoker:1999mqo} to make an appearance. One possible way to organize this calculation is to use the results of \cite{Prabhu:2020avf} and express such correlators in terms of the building blocks obtained here by studying scalar correlators.

The approach used here to treat correlators in flat space is very similar to the Mellin space approach to CFT correlators \cite{Penedones:2010ue,Fitzpatrick:2011ia}. This has been used to bootstrap CFTs in Mellin space \cite{Gopakumar:2016wkt,Gopakumar:2016cpb}. It would be very interesting if those ideas can be used in the context of the correlators presented here.

Perhaps, the most exciting question is about how much of the beautiful structure of Conformal Field Theories can be used to analyse massless field theories in flat spacetimes. Can we use the convergent OPE and well defined Hilbert space of the latter to our advantage? We hope to address such intriguing questions in the future.

\section*{Acknowledgments}
It is a pleasure to thank Abhijit Gadde,  Matt von Hippel, Diksha Jain, Alok Laddha,  R. Loganayagam, Suvrat Raju, Ashoke Sen, and especially Shiraz Minwalla and Onkar Parrikar for valuable discussions. Preliminary versions of this work were presented at the Niels Bohr Institute and at CEICO, and we are thankful for the discussions there. This work was supported by SERB, Department of Science and Technology under the grant PDF/2021/004777. We also acknowledge support from a J. C. Bose Fellowship JCB/2019/000052, and from the Infosys Endowment for the study of the Quantum Structure of Spacetime. Research at TIFR is supported by the Department of Atomic
Energy under Project Identification No. RTI4002. We would like to express our gratitude to the people of India for their support in the development of the basic sciences.

\appendix 
\section*{Appendix}
\section{Conformal contact contributions} \label{confcontact}
In this section, we shall find contact contributions to conformal correlators with  up to six external fields. Our starting point is the expression in \eqref{schcftn}, which, after introducing a Schwinger parameter to exponentiate the denominator, takes the form
\begin{equation} \label{schcftnexp}
	\begin{split}
		\gcfta[1,n][1,n]&=  \prod_{i=1}^n \int_{0}^{\infty} d\alpha_i \alpha_i^{\Delta_i-1} \delta(1-\sum\limits_{i=1}^n \lambda_i \alpha_i)  \int ds \, s^{d/2-1}\, e^{- s \sum_{i,j} \alpha_i \alpha_j s_{ij}}
	\end{split}
\end{equation}

\subsection{3 points} \label{cft3pt}
For the special and simple case of three points, we first take a different route and begin with the simplified expression in \eqref{cftnminusone} for the $(n+1)$-point conformal contact correlator, where one of the Schwinger parameter integrals is already done
\begin{equation} 
	\begin{split}
		\gcfta[1,n+1][1,n+1]=&\Gamma(\D[n]-d/2) \Gamma(d-\D[n])\\
		&\times \prod_{i=1}^{n} \int_{0}^{\infty} d\alpha_i  \alpha_i^{\Delta_i-1}  \delta(1-\sum\limits_{i=1}^{n} \lambda_i \alpha_i)  \frac{(\sum\limits_{i=1}^{n}\alpha_i s_{i (n+1)})^{\D[n]-d}}{\left(\sum\limits_{\substack{ i,j \neq n+1}} \alpha_i \alpha_j s_{ij}\right)^{\D[n]-d/2}}
	\end{split}
\end{equation}
Let us choose $\lambda_{n}=1$ and $\lambda_i=0$ for all other $i$. The Dirac delta function then localizes the $\alpha_n$ integral at $\alpha_n=1$ so that we get

\begin{equation} \label{nminus2}
	\begin{split}
		\gcfta[1,n+1][1,n+1]=&\Gamma(\D[n]-d/2) \Gamma(d-\D[n])\\
		&\times \prod_i^{n-1} \int_{0}^{\infty} d\alpha_i \alpha_i^{\Delta_i-1} \frac{(\sum\limits_{i=1}^{n-1}\alpha_i s_{i (n+1)}+s_{n(n+1)})^{\D[n]-d}}{(\sum\limits_{i=1}^{n-1}\alpha_i s_{i (n)}+\sum\limits_{\substack{ i,j \neq n,n+1}} \alpha_i \alpha_j s_{ij})^{\D[n]-d/2}}  
	\end{split}
\end{equation}
Evaluating \eqref{nminus2} for the simple case of three points gives

\begin{equation}
	\begin{split}
		\gcfta[1,3][1,3]&=\Gamma(\Delta_{12}-d/2) \Gamma(d-\Delta_{12})  \int_{0}^{\infty} d\alpha_1 \alpha_1^{\Delta_1-1} \frac{(\alpha_1 s_{1 3}+s_{23})^{\Delta_{12}-d}}{(\alpha_1 s_{12})^{\Delta_{12}-d/2}} 
	\end{split}
\end{equation}
The $\alpha_1$ integral is evaluated easily using \eqref{elemints}. Defining $\Delta_{ij,k}=\Delta_i+\Delta_j-\Delta_k$,  we find
\begin{equation} \label{cft3ptcont}
	\begin{split}
		\gcfta[1,3][1,3]
		&= \frac{\Gamma(\half \Delta_{12,3}) \Gamma(\half\Delta_{23,1}) \Gamma(\half \Delta_{13,2})}{\left(s_{12}\right)^{\half\Delta_{12,3}}(s_{23})^{\half\Delta_{23,1}}(s_{13})^{\half\Delta_{13,2}}} 
	\end{split}
\end{equation}
which is the familiar expression for the CFT 3-point function.

Anticipating its usefulness in the generalization to higher point correlators, we also compute the 3-point conformal contact contribution  starting from the expression in \eqref{schcftnexp},
\begin{equation} \label{schcft3exp}
	\begin{split}
		\gcfta[1,3][1,3]&=  \prod_{i=1}^3 \int_{0}^{\infty} d\alpha_i \alpha_i^{\Delta_i-1} \delta(1-\sum\limits_{i=1}^3 \lambda_i \alpha_i)  \int ds \, s^{d/2-1}\, e^{- s \sum_{i,j} \alpha_i \alpha_j s_{ij}}
	\end{split}
\end{equation}
Choosing $\lambda_3=1$ and $\lambda_1=\lambda_2=0$ localizes the $\alpha_3$ integral at $\alpha_3=1$ so that we get
\begin{equation} \label{schcft3exp}
	\begin{split}
		\gcfta[1,3][1,3]&=    \int_{0}^{\infty} d\alpha_1  \alpha_1^{\Delta_1-1} \int_{0}^{\infty} ds \, s^{d/2-1}  e^{-s \alpha_1 s_{13}} \int_{0}^{\infty} d\alpha_2  \alpha_2^{\Delta_2-1}\, e^{- s \, \alpha_2(s_{23}+\alpha_1 s_{12})}
	\end{split}
\end{equation}
It is easy to see, in this arrangement, that all the integrals are elementary and one of the following two kinds
\begin{equation} \label{elemints}
\begin{split}
		\int_0^{\infty} dx x^{a-1}  e^{-s x}&= \frac{\Gamma(a)}{s^{a}}\\
			\int_0^{\infty} dx x^{a-1} (1+ q x)^{-b} &=\frac{1}{q^{a}} \frac{\Gamma(a) \Gamma(b-a)}{\Gamma(b)} \\
\end{split}
\end{equation}
Performing all the integrals then and using $\D[3]=d$ precisely lands us at \eqref{cft3ptcont}.
\subsection{4 points} \label{cft4pt}
For the case $n=4$, the expression in \eqref{schcftnexp} becomes
\begin{equation} \label{schcft4exp}
	\begin{split}
		\gcfta[1,4][1,4]&=  \prod_{i=1}^4 \int_{0}^{\infty} d\alpha_i \alpha_i^{\Delta_i-1} \delta(1-\sum\limits_{i=1}^4 \lambda_i \alpha_i)  \int ds \, s^{d/2-1}\, e^{- s \sum_{i,j} \alpha_i \alpha_j s_{ij}}
	\end{split}
\end{equation}
Using the Dirac delta function to localize the $\alpha_4$ integral at $\alpha_4=1$, the exponent of the exponential is seen to be
\begin{equation} \label{exp4}
	\begin{split}
    &-s  \alpha_3 \left(s_{34}+\alpha_2 s_{23}+\alpha_1 s_{13} \right) \\
    &-s \alpha_2 (s_{24}+\alpha_1 s_{12}) \\
    &-s \alpha_1 s_{14} 
	\end{split}
\end{equation}
Let us look at the dependence of the above exponent on $\alpha_2$ and $\alpha_1$. From the computation of the 3-point correlator, we know that just the last two lines of the above exponent give rise to elementary integrals over $\alpha_2$ and $\alpha_1$. If we are able to convert the exponential dependence on $\alpha_2$ and $\alpha_1$ from the first line of \eqref{exp4} into power law dependencies on these two variables, then all the $\alpha$ integrals will turn out to be elementary. We can achieve this by use of the identity (as was noticed in \cite{Symanzik:1972wj}))
 \begin{equation}\label{expmellin}
	e^{-x}=\int_{c-i \infty}^{c+i \infty} {ds \over {2\pi i}} \Gamma(-s) x^{s}
\end{equation}
for the two exponentials $e^{-s  \alpha_3 \alpha_2 s_{23} }$ and $e^{-s  \alpha_3 \alpha_1 s_{13} }$ to arrive at
\begin{equation} 
	\begin{split}
		\gcfta[1,4][1,4]=&  \int_{c-i \infty}^{c+i \infty} {da_1 \over {2\pi i}}  \Gamma(-a_1) s_{13}^{a_1} \int_{c-i \infty}^{c+i \infty} {da_2 \over {2\pi i}} \Gamma(-a_2) s_{23}^{a_2}\\
		&\times \int_{0}^{\infty} d\alpha_1  \alpha_1^{\Delta_1+a_1-1} \int_{0}^{\infty} ds \, s^{d/2+a_{12}-1}  e^{-s \alpha_1 s_{14}} \int_{0}^{\infty} d\alpha_2  \alpha_2^{\Delta_2+a_2-1}\, e^{- s \alpha_2(s_{24}+\alpha_1 s_{12})}\\
		&  \times  \int_{0}^{\infty} d\alpha_3 \alpha_3^{\Delta_3+a_{12}-1} e^{- s \alpha_3  s_{34}}  
	\end{split}
\end{equation}
Again, it is seen that, when ordered in this manner, all the $\alpha$ integrals and the $s$ integral are elementary, and are either one of the two types in \eqref{elemints}. Performing all these integrals lands us at \eqref{gcft4}.
\subsection{5 points} \label{cft5pt}
Using \eqref{schcftnexp}, the conformal contact contribution for five external points can be seen to be
\begin{equation} \label{schcft5exp}
	\begin{split}
		\gcfta[1,5][1,5]&=  \prod_{i=1}^5 \int_{0}^{\infty} d\alpha_i \alpha_i^{\Delta_i-1} \delta(1-\sum\limits_{i=1}^5 \lambda_i \alpha_i)  \int ds \, s^{d/2-1}\, e^{- s \sum_{i,j} \alpha_i \alpha_j s_{ij}}
	\end{split}
\end{equation}
This time, we localize the $\alpha_5$ integral at $\alpha_5=1$, and find that the exponent of the exponential can be written as
\begin{equation} \label{exp5}
	\begin{split}
		&-s  \alpha_4 \left(s_{45}+\alpha_3 s_{34}+\alpha_2 s_{24} +\alpha_1 s_{14} \right) \\
		&-s  \alpha_3 \left(s_{35}+\alpha_2 s_{23}+\alpha_1 s_{13} \right) \\
		&-s \alpha_2 (s_{25}+\alpha_1 s_{12}) \\
		&-s \alpha_1 s_{15} 
	\end{split}
\end{equation}
Continuing with the flow of logic from the computation of the 3-point and 4-point correlators, we see that if we were to convert the exponentials coming from the last three terms in the first line of \eqref{exp5}, and those coming from the last two terms of the second line of \eqref{exp5} into powers using the identity \eqref{expmellin}, then all the $\alpha$ integrals and the $s$ integral turn out to be of the elementary kinds in \eqref{elemints}. The number of Mellin substitutions \eqref{expmellin} needed is $2+3=5$ which matches the number of conformal invariants with 5 points so that we have
\begin{equation} 
	\begin{split}
		\gcfta[1,5][1,5]&=  \prod_{i=1}^5 \int {da_i \over {2\pi i}}  \Gamma(-a_i)  \prod_{j=1}^3 s_{j4}^{a_j} \prod_{k=1}^2 s_{k3}^{a_{k+3}} \\
		& \times\int_{0}^{\infty} d\alpha_1  \alpha_1^{\Delta_1+a_{14}-1} \int_{0}^{\infty} ds \, s^{d/2+\A[5]-1}  e^{-s \alpha_1 s_{15}} \int_{0}^{\infty} d\alpha_2  \alpha_2^{\Delta_2+a_{25}-1}\, e^{- s \alpha_2(s_{25}+\alpha_1 s_{12})}\\
		&  \times \int_{0}^{\infty} d\alpha_3 \alpha_3^{\Delta_3+a_{345}-1} e^{- s \alpha_3  s_{35}}   \int_{0}^{\infty} d\alpha_4 \alpha_4^{\Delta_4+a_{123}-1} e^{- s \alpha_4  s_{45}} . 
	\end{split}
\end{equation}
Performing all the $\alpha$ integrals and the $s$ integral, we arrive at \eqref{gcft5}.

\subsection{6 points} \label{cft6pt}
Evaluating \eqref{schcftnexp} for $n=6$ by localizing the $\alpha_6$ integral at $\alpha_6=1$, the exponent of the exponential in the resulting expression is
\begin{equation} \label{exp6}
	\begin{split}
		&-s  \alpha_5 \left(s_{56}+\alpha_4 s_{45}+\alpha_3 s_{35}+\alpha_2 s_{25} +\alpha_1 s_{15} \right) \\
		&-s  \alpha_4 \left(s_{46}+\alpha_3 s_{34}+\alpha_2 s_{24} +\alpha_1 s_{14} \right) \\
		&-s  \alpha_3 \left(s_{36}+\alpha_2 s_{23}+\alpha_1 s_{13} \right) \\
		&-s \alpha_2 (s_{26}+\alpha_1 s_{12}) \\
		&-s \alpha_1 s_{16} 
	\end{split}
\end{equation}
This time, we see that we need to use the identity \eqref{expmellin} on the following exponentials: the last four terms of the first line, the last three terms of the second line and the last two terms of the third line. In other words, the number of new variables introduced is $2+3+4=9$. Here, we note that continuing this counting for the $n$-point correlator, the number of such variables is equal to $2+3+\ldots+(n-2)=n(n-3)/2$ which matches the counting of conformally invariant cross ratios that can be constructed from $n$ points.

We end up with the following expression for the six-point conformal contact correlator,
\begin{equation} 
	\begin{split}
		\gcfta[1,6][1,6]&=  \prod_{i=1}^9 \int {da_i \over {2\pi i}}  \Gamma(-a_i)  \prod_{j=1}^4 s_{j5}^{a_j} \prod_{k=1}^3 s_{k4}^{a_{k+4}} \prod_{\ell=1}^2 s_{\ell 3}^{a_{\ell+7}}\\
		& \times\int_{0}^{\infty} d\alpha_1  \alpha_1^{\Delta_1+a_{158}-1} \int_{0}^{\infty} ds \, s^{d/2+\A[9]-1}  e^{-s \alpha_1 s_{16}} \int_{0}^{\infty} d\alpha_2  \alpha_2^{\Delta_2+a_{269}-1}\, e^{- s \alpha_2(s_{26}+\alpha_1 s_{12})}\\
		& \times  \int_{0}^{\infty} d\alpha_3 \alpha_3^{\Delta_3+a_{3789}-1} e^{- s \alpha_3  s_{36}}  \int_{0}^{\infty} d\alpha_4 \alpha_4^{\Delta_4+a_{4567}-1} e^{- s \alpha_4  s_{46}}  \\
		&  \times \int_{0}^{\infty} d\alpha_5 \alpha_5^{\Delta_5+a_{1234}-1} e^{- s \alpha_5  s_{56}}  
	\end{split}
\end{equation}
We perform all the $\alpha$ integrals and the $s$ integral all of which are one of the two types in \eqref{elemints}, and find that the 6-point function can be expressed as
\begin{equation} \label{gcft6}
		\gcfta[1,6][1,6]=\left(\frac{s_{12}}{s_{16}s_{26}}\right)^{\Delta_6-d/2}  \frac{1}{\prod_{i=1}^{5}s_{i6}^{\Delta_i}}\dcft_{\Delta_1,\ldots,\Delta_6}(s_1,\ldots,s_9)
\end{equation}
with 
\begin{equation} \label{dcft6}
	\begin{split}
		\dcft_{\Delta_1,\ldots,\Delta_6}(s_1,\ldots,s_9)
		= &\prod_{i=1}^{9}\int  {da_i \over {2\pi i}}   \,s_i^{a_i}  \,\Gamma(-a_i)  \Gamma(\Delta_5+a_{1234})\Gamma(\Delta_4+a_{4567})  \Gamma(\Delta_3+a_{3789}) \\
		&\times \Gamma\left(\Delta_{16}-{d \over2}-a_{234679}\right) \Gamma\left(\Delta_{26}-{d \over2}-a_{134578}\right) \Gamma\left({d \over2}-\Delta_6-\A[9]\right)
	\end{split}	
\end{equation}
a function of the cross ratios 
\begin{equation} \label{cr6}
	\begin{split}
		&s_1=\frac{s_{15}s_{26}}{s_{12}s_{56}}, s_2=\frac{s_{25}s_{16}}{s_{12}s_{56}},s_3=\frac{s_{35}s_{16}s_{26}}{s_{12}s_{36}s_{56}},s_4=\frac{s_{45}s_{16}s_{26}}{s_{12}s_{46}s_{56}},\\
		&s_5=\frac{s_{14}s_{26}}{s_{12}s_{46}},	s_6=\frac{s_{24}s_{16}}{s_{12}s_{46}}, s_7=\frac{s_{34}s_{16}s_{26}}{s_{12}s_{36}s_{46}},\\
		&s_8=\frac{s_{13}s_{26}}{s_{12}s_{36}},s_9=\frac{s_{23}s_{16}}{s_{36}s_{12}}
	\end{split}
\end{equation}

\section{Conformal exchanges}
In \S \ref{exch}, we computed the 1-exchange diagram contribution to the $n-$point massless correlator. We can find the 1-exchange contribution to the $n-$point conformal correlator by  using conformality constraints on the scaling dimensions $\Delta_L=\Delta_R=d$  in \eqref{flat1e} as

\begin{equation} \label{cft1e}
	{\gcft}_{e,n} =   \prod_{i=0}^{n} \int_{0}^{\infty} da_i a_i^{\Delta_i-1}e^{- \sum\limits_{ L' } a_i a_j s_{ij}}  e^{- \sum\limits_{L'+R'} a'_i a'_j s_{ij}}
\end{equation}
with the definitions
\begin{equation}\label{aprimep}
	a'_i = \begin{cases}
		a_0 a_i & i\in L' \\
		a_i & i \in R'
	\end{cases}
\end{equation}
and $A_L'=\sum\limits_{i\in L'} a_i$.  Here $L'=\{1,\ldots,\ell\}$ and $R'=\{\ell+1,\ldots,n\}$.
We would like to do the $a_0$ integral first, and attempt to do the rest of the integrals in a fashion similar to the contact case. To this end, we make the variable substitution
\begin{equation}\label{ap}
	a_i = \begin{cases}
		t_0 & i =0\\
		{t_i \over t_0} & i\in L' \\
		t_i & i \in R'
	\end{cases}
\end{equation}
to find that
\begin{equation} 
	{\gcft}_{e,n} =   \prod_{i=1}^{n} \int_{0}^{\infty} dt_i t_i^{\Delta_i-1} e^{- \sum\limits_{L'+R'} t_i t_j s_{ij}}  \int_{0}^{\infty} dt_0 \,t_0^{\Delta_0-\Delta_L-1} e^{- \frac{1}{t_0^2}\sum\limits_{ L' } t_i t_j s_{ij}}  
\end{equation}
Using a slightly modified version of our familiar trick again, we introduce $$ 1= \int_0^{\infty} ds \delta(\sqrt{s}-\sum\limits_{i=1}^{n} \lambda_i t_i) $$ for arbitrary $\lambda_i \geq 0$, and then scale $t_i \rightarrow \sqrt{s} \alpha_i$ and $t_0 \rightarrow \sqrt{s} \alpha_0$ to find 
\begin{equation} 
	\begin{split}
		{\gcft}_{e,n} =   \prod_{i=1}^{n} \int_{0}^{\infty} d\alpha_i \alpha_i^{\Delta_i-1} \delta(1-\sum_{i=1}^n \lambda_i \alpha_i) \int_{0}^{\infty}d\alpha_0 \,\alpha_0^{\Delta_0-d-1} e^{- \frac{1}{\alpha_0^2}\sum\limits_{ L' } \alpha_i \alpha_j s_{ij}}  \int ds s^{d/2-1} e^{- s\sum\limits_{L'+R'} \alpha_i \alpha_j s_{ij}} 
	\end{split}
\end{equation}
where we used the conformality constraints $\Delta_L=\Delta_R=d$. Now performing the elementary integrals in $s$ and $\alpha_0$, we obtain
\begin{equation} \label{cft1edenom}
	\begin{split}
		{\gcft}_{e,n} =   \prod_{i=1}^{n} \int_{0}^{\infty} d\alpha_i \alpha_i^{\Delta_i-1} \delta(1-\sum_{i=1}^n \lambda_i \alpha_i) \frac{1}{(\sum\limits_{ L' } \alpha_i \alpha_j s_{ij})^{d/2-\Delta_0}}  \frac{1}{(\sum\limits_{ L' +R'} \alpha_i \alpha_j s_{ij})^{d/2}}  
	\end{split}
\end{equation}
This integral is not too much more complicated than the corresponding integral for the contact contribution. This can be seen by exponentiating the two denominators by introducing two additional Schwinger parameters, and then noting that the number of substitutions of the kind in \eqref{expmellin} required for the exponential functions is equal to the number of cross ratios, and so, is same as that for the corresponding contact contribution. We carry this exercise for the first non-trivial case of six external fields (1-exchange diagram with two 4-point vertices) in Appendix \ref{1exchexact}. This can be used to obtain the answer for the non-conformal 1-exchange contribution to the 4-point correlator.
\subsection{1-exchange diagram for the six-point correlator} \label{1exchexact}
In this section, we will compute the exact answer for the conformal 1-exchange diagram in Fig. \ref{6pt1exch} with two four point vertices given by
\begin{equation} \label{gcfte6}
	{\gcft}_{e,6} ={1 \over \pi^d}\int d^d y_1 \left(\prod_{i=1}^{3}\frac{\Gamma(\d[i])}{((x_i-y_1)^2)^{\Delta_i}}\right) \int d^d y_2 \left(\prod_{i=4}^{6}\frac{\Gamma(\d[i])}{((x_i-y_2)^2)^{\Delta_i}} \right) \frac{1}{((y_1-y_2)^2)^{\Delta_0}}   
\end{equation}

Specializing our answer for the 1-exchange diagram with any number of external fields in \eqref{cft1edenom} to the case of six external fields, we find that
\begin{equation} 
	\begin{split}
		{\gcft}_{e,6}
		&=   \prod_{i=1}^{6} \int_{0}^{\infty} d\alpha_i \alpha_i^{\Delta_i-1} \delta(1-\sum_{i=1}^6 \lambda_i \alpha_i) \int dp \, p^{d/2-\Delta_0-1}e^{-p \sum\limits_{ L' } \alpha_i \alpha_j s_{ij}} \int ds \, s^{d/2-1}e^{-s \sum\limits_{ L'+R' } \alpha_i \alpha_j s_{ij}}
	\end{split}	
\end{equation}

Combining the two exponentials, the resultant exponent is seen to be
\begin{equation} \label{exp1exch}
	\begin{split}
		&-  \alpha_5 \left\{\left(p+s\right)\left(s_{56}+\alpha_4 s_{45}\right)+s\left(\alpha_3 s_{35}+\alpha_2 s_{25} +\alpha_1 s_{15} \right)\right\} \\
		&- \alpha_4 \left\{\left(p+s\right)s_{46}+s\left(\alpha_3 s_{34}+\alpha_2 s_{24} +\alpha_1 s_{14} \right) \right\}\\
		&-s  \alpha_3 \left(s_{36}+\alpha_2 s_{23}+\alpha_1 s_{13} \right) \\
		&-s \alpha_2 (s_{26}+\alpha_1 s_{12}) \\
		&-s \alpha_1 s_{16} 
	\end{split}
\end{equation}
Again, we use the identity \eqref{expmellin} on the last four terms of the first line, the last three terms of the second, and the last two terms of the third line, to find 
\begin{equation} 
	\begin{split}
		{\gcft}_{e,6}=&  \prod_{i=1}^9 \int {da_i \over {2\pi i}}  \Gamma(-a_i)  \prod_{j=1}^4 s_{j5}^{a_j} \prod_{k=1}^3 s_{k4}^{a_{k+4}} \prod_{\ell=1}^2 s_{\ell 3}^{a_{\ell+7}}\\
		& \times\int_{0}^{\infty} d\alpha_1  \alpha_1^{\Delta_1+a_{158}-1} \int_{0}^{\infty} ds \, s^{d/2+\A[9]-a_4-1}  e^{-s \alpha_1 s_{16}} \int_{0}^{\infty} d\alpha_2  \alpha_2^{\Delta_2+a_{269}-1}\, e^{- s \alpha_2(s_{26}+\alpha_1 s_{12})}\\
		&  \times \int_{0}^{\infty} d\alpha_3 \alpha_3^{\Delta_3+a_{3789}-1} e^{- s \alpha_3  s_{36}}   \times\int_{0}^{\infty} dp \,p^{d/2-\d[0]-1}  (p+s)^{p_4}\\
		&  \times \int_{0}^{\infty} d\alpha_4 \alpha_4^{\Delta_4+a_{4567}-1} e^{- (p+s) \alpha_4  s_{46}}  \int_{0}^{\infty} d\alpha_5 \alpha_5^{\Delta_5+a_{1234}-1} e^{- (p+s) \alpha_5  s_{56}}  
	\end{split}
\end{equation}
As usual, using \eqref{elemints} to do all the $\alpha$ and $s$ integrals, we obtain the 1-exchange contribution to the six-point conformal correlator as
\begin{equation} \label{gcfte6}
	{\gcft}_{e,6} =\Gamma(d/2-\Delta_0)\left(\frac{s_{16}s_{26}}{s_{12}}\right)^{\Delta_{123}-\Delta_6}  \frac{1}{\prod_{i=1}^{5}s_{i6}^{\Delta_i}}   \dcft_{e,6}(s_1,\ldots,s_9)
\end{equation}
with 
\begin{equation} \label{dcfte6}
	\begin{split}
		\dcft_{e,6}(s_1,\ldots,s_9)
		= &\prod_{i=1}^{9}\int  {da_i \over {2\pi i}}   \,s_i^{a_i}  \,\Gamma(-a_i)  \Gamma(\Delta_5+a_{1234})\Gamma(\Delta_4+a_{4567})  \Gamma(\Delta_3+a_{3789})\Gamma\left(\Delta_{123}-\Delta_6+\A[9]\right) \\
		&\times \Gamma\left(\Delta_{6}-\Delta_{23}-a_{234679}\right) \Gamma\left(\Delta_{6}-\Delta_{13}-a_{134578}\right) \frac{\Gamma(d/2-\Delta_6+\A[7])}{\Gamma(\Delta_{45}+\A[7])}
	\end{split}	
\end{equation}
a function of the cross ratios given in \ref{cr6}

\bibliographystyle{JHEP.bst}
\bibliography{references.bib}

\end{document}